\documentclass{eptcs}

\newif\ifignore 
\ignorefalse
\newcommand{\auxproof}[1]{
\ifignore\mbox{}\newline
\textbf{PROOF:} \dotfill\newline
{\it #1}\mbox{}\newline
\textbf{ENDPROOF}\dotfill
\fi}

\usepackage{graphicx}
\usepackage{etex} 
\usepackage{amsmath}
\usepackage{amssymb}
\usepackage{amstext}
\usepackage{amsthm}
\usepackage{mathtools}
\usepackage{mathrsfs}
\usepackage{stmaryrd}
\usepackage{xspace}
\usepackage{enumitem}
\usepackage{cmll}
\usepackage{url}
\usepackage{nicefrac}
\usepackage{listings}

\usepackage{xypic}
\usepackage[all,cmtip]{xy}
\xyoption{tips}

\usepackage{tikzit}
\usetikzlibrary{circuits.ee.IEC}
\usetikzlibrary{shapes,shapes.geometric,shapes.misc}

\tikzstyle{white dot}=[inner sep=0mm, minimum size=1.5mm, draw=black, shape=circle, text depth=-0.2mm, draw=black, fill=white, tikzit category=nodes]
\tikzstyle{black dot}=[inner sep=0mm, minimum size=1.5mm, draw=black, shape=circle, draw=black, fill=black, tikzit category=nodes]
\tikzstyle{observed}=[inner sep=0mm, minimum size=5mm, draw=black, shape=circle, text depth=-0.2mm, draw=white, tikzit draw=gray, fill=white, tikzit category=dag]
\tikzstyle{latent}=[inner sep=0mm, minimum size=5mm, draw=black, shape=circle, text depth=-0.2mm, draw=black, fill=white, tikzit category=dag]
\tikzstyle{small box}=[shape=rectangle, text height=1.5ex, text depth=0.25ex, yshift=0.5mm, fill=white, draw=black, minimum height=6mm, yshift=-0.5mm, minimum width=6mm, font={\small}, tikzit category=boxes]
\tikzstyle{medium box}=[shape=rectangle, draw=black, fill=white, small box, minimum width=8mm, tikzit category=boxes]
\tikzstyle{semilarge box}=[shape=rectangle, draw=black, fill=white, small box, minimum width=12.5mm, tikzit category=boxes]
\tikzstyle{large box}=[shape=rectangle, draw=black, fill=white, small box, minimum width=15mm, tikzit category=boxes]
\tikzstyle{upground}=[circuit ee IEC, thick, ground, rotate=90, scale=1.5, inner sep=-2mm, tikzit shape=circle, tikzit fill=blue, tikzit category=points]
\tikzstyle{downground}=[circuit ee IEC, thick, ground, rotate=-90, scale=1.5, inner sep=-2mm, tikzit shape=circle, tikzit fill=green, tikzit category=points]
\tikzstyle{point}=[regular polygon, regular polygon sides=3, draw, scale=0.75, inner sep=-0.5pt, minimum width=9mm, fill=white, regular polygon rotate=180, tikzit category=points]
\tikzstyle{copoint}=[regular polygon, regular polygon sides=3, draw, scale=0.75, inner sep=-0.5pt, minimum width=9mm, fill=white, tikzit category=points]
\tikzstyle{uniform}=[point, fill=gray, tikzit shape=circle, scale=0.5]
\tikzstyle{label}=[font={\footnotesize}, text height=1.5ex, text depth=0.25ex, tikzit draw=blue, tikzit fill=white, tikzit category=labels]
\tikzstyle{left label}=[label, anchor=east, xshift=2mm, tikzit draw=green, tikzit fill=white, tikzit category=labels]
\tikzstyle{right label}=[label, anchor=west, xshift=-2mm, tikzit draw=purple, tikzit fill=white, tikzit category=labels]
\tikzstyle{disintegration}=[draw=black, fill={gray!50}, tikzit fill=gray, shape=rectangle, minimum width=1.6cm, minimum height=1.2cm, opacity=0.3]
\tikzstyle{empty diag}=[shape=rectangle, draw=darkgray, dashed, minimum width=8mm, minimum height=8mm, yshift=0.5mm]

\tikzstyle{diredge}=[->, >=latex]
\tikzstyle{dashed edge}=[-, dashed]

\ifdefined\mathsl\else
  \DeclareMathAlphabet{\mathsl}{\encodingdefault}{\rmdefault}{\mddefault}{\sldefault}
  \SetMathAlphabet{\mathsl}{bold}{\encodingdefault}{\rmdefault}{\bfdefault}{\sldefault}
\fi

\newenvironment{myproof}{\begin{trivlist} \item[\hskip \labelsep%
{\bf Proof.}]}{\end{trivlist}}

\newcommand{\QEDbox}{\square}
\newcommand{\QED}{\hspace*{\fill}$\QEDbox$}

\makeatletter
\newcommand{\mathoverlap}[2]{\mathpalette\mathoverlap@{{#1}{#2}}}
\newcommand{\mathoverlap@}[2]{\mathoverlap@@{#1}#2}
\newcommand{\mathoverlap@@}[3]{\ooalign{$\m@th#1#2$\crcr\hidewidth$\m@th#1#3$\hidewidth}}
\makeatother

\newcommand{\pull}{\mathrel{\mathchoice%
   {\scalebox{-0.5}[1]{$\gg=$}}
   {\scalebox{-0.5}[1]{$\gg{\kern-1.5ex}=$}}
   {\scalebox{-0.5}[1]{${\kern.5ex}\scriptstyle\gg{\kern-0.2ex}={\kern.5ex}$}}
   {\scalebox{-0.5}[1]{$\scriptscriptstyle\gg=$}}}}
\newcommand{\push}{\mathrel{\mathchoice%
   {\scalebox{-0.5}[1]{$=\ll$}}
   {\scalebox{-0.5}[1]{$={\kern-1.5ex}\ll$}}
   {\scalebox{-0.5}[1]{${\kern.5ex}\scriptstyle={\kern-0.2ex}\ll{\kern.5ex}$}}
   {\scalebox{-0.5}[1]{$\scriptscriptstyle=\ll$}}}}

\DeclareSymbolFont{T1op}{T1}{cmr}{m}{n}
\SetSymbolFont{T1op}{bold}{T1}{cmr}{bx}{n}
\DeclareMathSymbol{\mathguilsinglleft}{\mathopen}{T1op}{'016}
\DeclareMathSymbol{\mathguilsinglright}{\mathclose}{T1op}{'017}
\newcommand{\klin}[1]{\mathguilsinglleft#1\mathguilsinglright}

\newcommand{\KLD}{\ensuremath{\mathsl{D}_{\mathsl{KL}}}}

\newcommand{\set}[2]{\{#1\;|\;#2\}}

\newcommand{\ket}[1]{\ensuremath{|{\kern.1em}#1{\kern.1em}\rangle}}
\newcommand{\bigket}[1]{\ensuremath{\big|{\kern.1em}#1{\kern.1em}\big\rangle}}

\newcommand{\one}{\ensuremath{\mathbf{1}}}

\newcommand{\andthen}{\mathrel{\&}}

\newcommand{\incr}[2]{#1\!\mathrel{\mathrm{+{\kern-.1em}+}}\!#2}

\newcommand{\distributionsymbol}{\mathcal{D}}

\newcommand{\Dst}{\distributionsymbol}

\newcommand{\UF}{\ensuremath{\mathcal{U}{\kern-.75ex}\mathcal{F}}}

\newcommand{\Kl}{\mathcal{K}{\kern-.4ex}\ell}

\newcommand{\EM}{\mathcal{E}{\kern-.4ex}\mathcal{M}}

\newcommand{\Pred}{\ensuremath{\mathrm{Pred}}}

\newcommand{\Ef}{\ensuremath{\mathcal{E}{\kern-.5ex}f}}
\newcommand{\intd}{{\kern.2em}\mathrm{d}{\kern.03em}}
\newcommand{\indic}[1]{\mathbf{1}_{#1}}

\newcommand{\OF}{\ensuremath{\mathcal{O}{\kern-.1em}\mathcal{F}}}
\newcommand{\Closed}{\ensuremath{\mathcal{C}{\kern-.05em}\ell}}

\newsavebox\sbpto
\savebox\sbpto{\begin{tikzpicture}[baseline=-2.4pt]
            \filldraw[fill=white,draw=white] circle (1.4pt);
            \filldraw[fill=white,draw=black,line width=0.2pt] circle (2.0pt);
                \end{tikzpicture}}
\newcommand\chanto{\mathrel{\ooalign{$\rightarrow$\cr
            \hfil\!$\usebox\sbpto$\hfil\cr}}}

\makeatother
 \newdir{ >}{{}*!/-7.5pt/@{>}}
 \newdir{|>}{!/4.5pt/@{|}*:(1,-.2)@^{>}*:(1,+.2)@_{>}}
 \newdir{ |>}{{}*!/-3pt/@{|}*!/-7.5pt/:(1,-.2)@^{>}*!/-7.5pt/:(1,+.2)@_{>}}

\newsavebox\sbground
\savebox\sbground{\begin{tikzpicture}[circuit ee IEC,yscale=0.5,xscale=0.4]
                \draw (0,-2ex) to (0,0) node[ground,rotate=90,xshift=.65ex] {};
                \end{tikzpicture}}

\newcommand{\eg}{\textit{e.g.}\xspace}

\theoremstyle{plain}
\newtheorem{theorem}{Theorem}
\newtheorem{lemma}{Lemma}
\newtheorem{proposition}[lemma]{Proposition}

\theoremstyle{definition}

\newtheorem{example}{Example}
\newtheorem{remark}{Remark}

\providecommand{\event}{MFPS 2021}
\title{Learning from What's Right and \\
       Learning from What's Wrong}

\author{Bart Jacobs\institute{Institute
    for Computing and Information Sciences (iCIS), 
\\ Radboud  University Nijmegen, The Netherlands.
}  
\email{bart@cs.ru.nl} 
}

\def\titlerunning{Learning from What's Right / Wrong}
\def\authorrunning{B. Jacobs}

\date{\small \today}

\begin{document}
\maketitle

\begin{abstract} 
The concept of updating (or conditioning or revising) a probability
distribution is fundamental in (machine) learning and in predictive
coding theory. The two main approaches for doing so are called Pearl's
rule and Jeffrey's rule. Here we make, for the first time,
mathematically precise what distinguishes them: Pearl's rule increases
validity (expected value) and Jeffrey's rule decreases
(Kullback-Leibler) divergence. This forms an instance of a more
general distinction between learning from what's right and learning
from what's wrong.  The difference between these two approaches is
illustrated in a mock cognitive scenario.
\end{abstract}


\section{Introduction}\label{IntroSec}

Intuitively, people can learn by reinforcing what goes well, or by
steering away from what goes wrong: they can go even higher, or go
lower. In the first case one improves a positive evaluation and in the
second case one reduces a negative outcome. In this paper we shall
refer to the first approach as learning from what's right, or more
simply, from rightness. The second approach is described in terms of
learning from what's wrong, or from wrongness.

Learning is at the heart of the current AI-revolution. In a
mathematical setting learning involves adapting/updating parameters,
with respect to some objective (expressed as `objective'
function). Also in such a setting one can distinguish whether this
adaptation is guided by increasing what is right, or by decreasing
what is wrong. Learning from rightness can be done by increasing a
positive evaluation, like reward, match, likelihood or
validity. Learning from wrongness happens by decreasing a negative
evaluation, like an error, loss, penalty, divergence or distance.
This distinction between learning from rightness/wrongness is not new
and may be expressed alternatively for instance in terms of
reward/error-based learning.  Also, the distinction is not absolute,
since what's good in one context may be bad in another.


In probabilistic learning there are two different approaches to
updating, namely following Pearl~\cite{Pearl90} (and Bayes) or
following Jeffrey~\cite{Jeffrey83}, see for comparisons
\textit{e.g.}~\cite{ChanD05,MradDPLA15,DietrichLB16,Jacobs19c}. The
two approaches can give completely different outcomes, but it is
poorly understood when to use which approach. For instance, it is
suggested by~\cite{DietrichLB16} that Jeffrey's rule is most
appropriate for correction after a `surprise', but it remains vague
what a surprise is. At a conceptual level, the main contribution of
this paper lies in showing that Pearl's approach is learning from
rightness, and Jeffrey's approach is learning from wrongness. The
objective function that is used here for rightness is validity (that
is, expected value), and for wrongness it is divergence (in
Kullback-Leibler form). Thus, it will be shown that Pearl's update
rule increases validity and Jeffrey's rule decreases divergence.  The
latter divergence-decrease result is the main mathematical
contribution of this paper.  Its proof makes use of rather heavy
mathematical machinery, taken mainly from~\cite{FriedlandK75}; it is
relegated to the appendix. The fact that Pearl's approach increases
validity is mathematically less complicated and the relevant parts of
this claim have already been published
\textit{e.g.}~in~\cite{Jacobs19b,JacobsZ21}.

Probabilistic learning typically involves an iterative process where
each single step yields an improvement w.r.t.\ an objective function.
This paper concentrates on these single steps, since its focus is on
capturing the (mathematical) difference between Pearl and Jeffrey.
What happens when these single learning steps are iterated is a topic
in itself, which is not covered here.

This paper builds on the mathematical formalisations of the rules of
Jeffrey and Pearl introduced in~\cite{Jacobs19c} --- where notably
Jeffrey's rule is captured via a `dagger'. In fact, had the main
result of this paper (Jeffrey reduces divergence) been known at the
time of writing~\cite{Jacobs19c}, it would have fitted perfectly
there. Alas, insights come slowly, and so a separate paper is written
now, as an addendum to~\cite{Jacobs19c}. The current addendum is much
more mathematical in nature than~\cite{Jacobs19c}, since the proof of
our main result is non-trivial. In addition, this addendum explains
the results in a more cognition-oriented language.

We thus start from the mathematical formalisation of~\cite{Jacobs19c}
that uses (discrete, finite) probability distributions (also called
states), fuzzy (soft) predicates, and channels (conditional
distributions), together with operations such as state/predicate
transformation along a channel and updating a distribution with a
predicate. Within this framework the update rules of Pearl and Jeffrey
are applied in a common setting, which we briefly introduce, without
explaining all details yet. One starts from a distribution $\sigma$ on
some set $X$ together with a channel $c$ from $X$ to $Y$, that is,
with a conditional probability distribution $p(y\mid x)$, or, more
categorically, with a Kleisli map of the distribution monad
$\Dst$. Along this channel one can transform (push forward) the
distribution $\sigma$ on $X$ to a distribution $c \push \sigma$ on $Y$,
namely $(c \push \sigma)(y) = \sum_{x} \sigma(x)\cdot p(y\mid x)$. This
$c \push \sigma$ can be seen as a prediction. We consider the situation
where we are confronted with new information (evidence) on $Y$ which
leads us to update $\sigma$ to a new distribution $\sigma'$. Pearl and
Jeffrey provide two different rules for performing this update, which
increase validity and decrease divergence, respectively. (In this
paper we only consider updating the state $\sigma$ along a channel
$c$, but one may go further and update the mediating channel $c$ as
well, like in Expectation Maximisation, see \cite{Jacobs19b}).  This
validity-increase and divergence-decrease is characteristic for the
rules of Pearl and Jeffrey: we demonstrate that Jeffrey's rule, in
general, does not give a validity increase, and similarly, that
Pearl's rule need not give a divergence-decrease (see
Remark~\ref{ExclusivityRem}).

Interestingly, the channel-based setting fits the neuroscientific
setting that underlies predictive coding theory (also called
predictive processing or free energy principle). This theory goes back
to Hermann von Helmholtz in the 19th century and is described in
modern terms first by~\cite{RaoB99} and in many other recent sources,
\textit{e.g.}~\cite{Friston10,Hohwy13,Clark16}. Naively, humans learn
by absorbing sensory information from the outside world and by
building up a more or less accurate internal picture. Alternatively,
the mind projects, evaluates and updates: predictive coding theory
describes the human mind basically as a Bayesian prediction engine
that compares its predictions to observations, leading to internal
adaptations.  To quote Friston~\cite{Friston10}: ``The \emph{Bayesian
  brain hypothesis} uses Bayesian probability theory to formulate
perception as a constructive process based on internal or generative
models. [\ldots] In this view, the brain is an inference machine that
actively predicts and explains its sensations. Central to this
hypothesis is a probabilistic model that can generate predictions,
against which sensory samples are tested to update beliefs about their
causes.'' We translate this to the above setting: the mind's internal
state may be (partially) represented by a distribution $\sigma$ on
$X$, as used in the previous paragraph. The channel $c$ is part of the
generative model that produces the prediction $c \push \sigma$, as
distribution on the external world $Y$. Confronted with (mismatching)
sensory information (about $Y$), the brain updates its internal state
$\sigma$ (on $X$). This is how learning happens in the predictive
model. This paper uses a running example of this kind.

An intriguiging question is: does this learning/updating happen
according to Pearl or to Jeffrey?  Formulated more abstractly, does
the mind learn from what's right or from what's wrong?  An (empirical)
answer to that question lies far beyond this paper, but predictive
coding theory suggests that our minds use Jeffrey's rule since they
try to minimise prediction errors, see \cite{Friston10}. An additional
argument in this direction is that successive Pearl-updates commute,
but successive Jeffrey-updates do not, see \cite{Jacobs19c} for
details. It is well-known that the human mind is highly sensitive to
the order in which it processes information (or: is primed/updated).
The main result of this paper (Theorem~\ref{JeffreyRuleThm})
strengthens the mathematical basis of predictive coding theory: it
extends learning from point data to learning from distributions and
shows that in such learning from distributions the prediction error,
expressed as Kullback-Leibler divergence, is reduced.

The structure of this paper is simple: after introducing preliminaries
in Section~\ref{PredictionSec}, Pearl's and Jeffrey's update rules are
described in Sections~\ref{PearlSec} and~\ref{JeffreySec},
following~\cite{Jacobs19c}. Section~\ref{JeffreyDecreaseSec} contains
the main new result of this paper, namely that Jeffrey's rule
decreases divergence. Its relevance to predictive coding theory is
explained in Section~\ref{PredCodSec}. Finally, the appendix contains
a proof of the main result.

\section{Prelimaries on states, channels and prediction}\label{PredictionSec}

This section introduces basic concepts and fixes notation.  Suppose
you mix paint of different colours, say with ratio $\frac{1}{2}$ red,
$\frac{1}{8}$ green and $\frac{3}{8}$ blue. In that case we can write
the paint distribution as a formal sum:
\[ \textstyle\frac{1}{2}\ket{R} + \frac{1}{8}\ket{G} + \frac{3}{8}\ket{B}. \]

\noindent The letters represent the different colours; they are
written between `ket' brackets $\ket{-}$ which are borrowed from
quantum physics. The kets are meaningless notation that serve to
separate the items in the distribution from their frequencies (or
probabilities).

In general a (finite, discrete) \emph{probability distribution} over a
set $X$ is a finite formal sum of the form $\sum_{i}r_{i}\ket{x_i}$
where the $r_{i}\in [0,1]$ are probabilities that add up to one:
$\sum_{i}r_{i} = 1$. The $x_i$ are members of the set $X$. Such a
distribution can also be written as a function $\omega\colon X
\rightarrow [0,1]$ with finite support, that is with only finitely
many $x\in X$ with $\omega(x) \neq 0$. We then have $r_{i} =
\omega(x_{i})$. We say that $\omega$ has \emph{full support} when
$\omega(x) > 0$ for each $x\in X$ --- which implicitly requires that
the set $X$ is finite. We use `state' as synonym for `distribution',
but the term `multinomial' is also common in the literature. We freely
switch between the formal sum notation $\omega =
\sum_{i}r_{i}\ket{x_i}$ and the function notation $\omega\colon X
\rightarrow [0,1]$. We write $\Dst(X)$ for the set of distributions on
the set $X$. This $\Dst$ is the (finite, discrete) distribution monad
on the category of sets.

A \emph{channel} is a Kleisli map for this distribution monad, that
is, a function of the form $c\colon X \rightarrow \Dst(Y)$. We say
that $c$ is a channel from $X$ to $Y$ and often write this as $c\colon
X \chanto Y$, with a small circle on the shaft of the arrow. Such a
channel gives a distribution $c(x)$ on $Y$ for each $x\in X$. It is
thus a conditional distribution, which is commonly written as $p(y\mid
x)$. Alternatively, when $X,Y$ are finite, we can see the channel as a
stochastic matrix. Each function $f\colon X \rightarrow Y$ gives rise
to a `deterministic' channel $\klin{f} \colon X \chanto Y$, via
$\klin{f}(x) = 1\ket{f(x)}$. Channels have a lot of algebraic
structure: they can be composed sequentially and also in parallel ---
they form the morphisms of a symmetric monoidal (Kleisli) category,
see \textit{e.g.}~\cite{Jacobs18c,Jacobs15d}. Channels are becoming
popular in a principled, axiomatic approach to probability, see
\textit{e.g.}~\cite{Fritz20,JacobsZ21,Jacobs21a,Jacobs21d}.

Let $c \colon X \chanto Y$ be a channel from $X$ to $Y$. We can then
define \emph{state tranformation} along $c$ as a function $\Dst(X)
\rightarrow \Dst(Y)$. It maps a distribution $\sigma$ on $X$ to a
`transformed' distribtution $c \push \sigma$ on $Y$, via the following
definition: for $y\in Y$,
\[ \begin{array}{rcl}
\big(c \push \sigma\big)(y)
& = &
\displaystyle\sum_{x\in X}\, \sigma(x)\cdot c(x)(y).
\end{array} \]

\noindent This new distribution is often called the \emph{prediction}.
This will be illustrated next in our leading example.

\begin{figure}
\begin{center}
\begin{tabular}{ccccc}
\includegraphics[width=4cm]{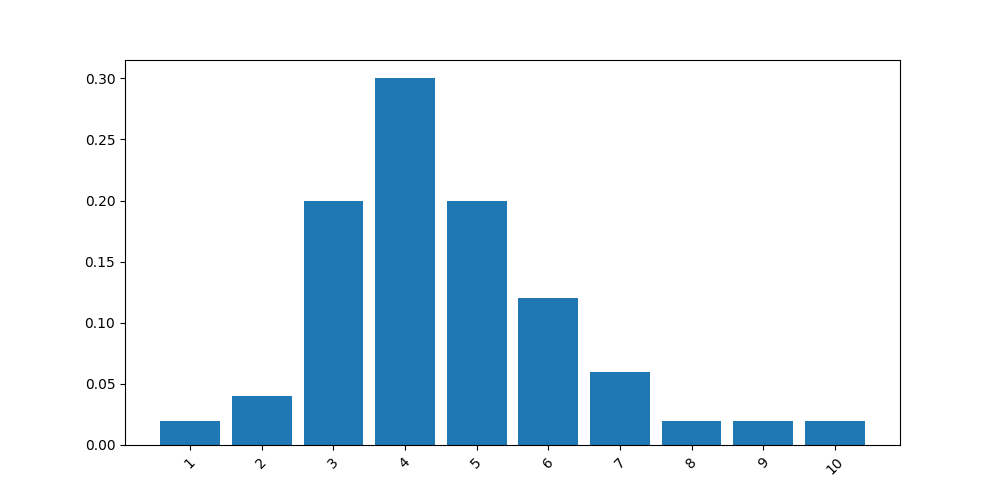}
& \quad &
\includegraphics[width=4cm]{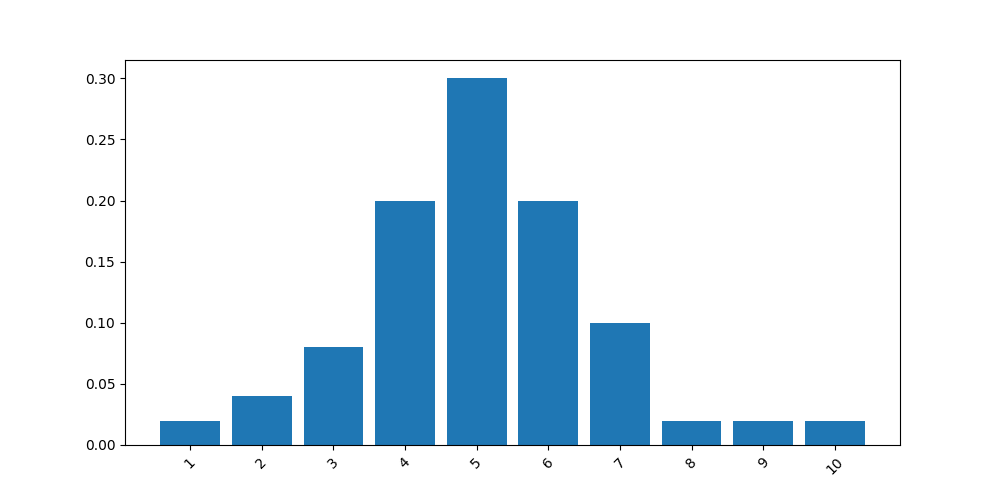}
& \quad &
\includegraphics[width=4cm]{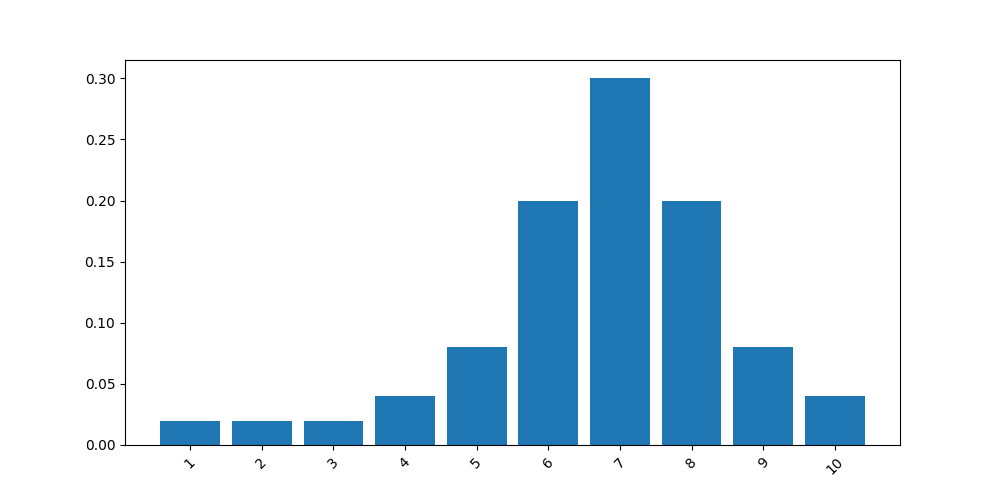}
\\[-0.7em]
pessimistic mood marks & &
neutral mood marks & &
optimistic mood marks
\end{tabular}
\\[+1em]
\begin{tabular}{ccc}
\includegraphics[width=4cm]{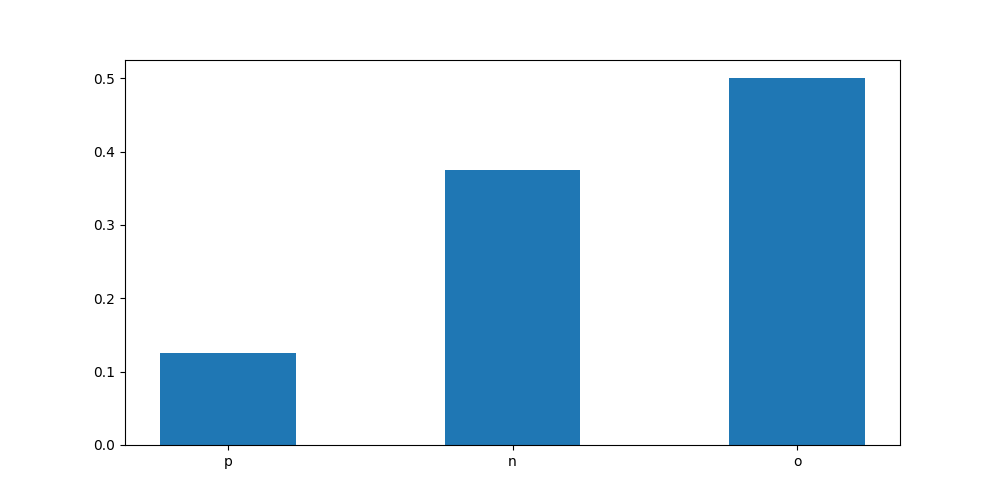}
& \qquad & 
\includegraphics[width=4cm]{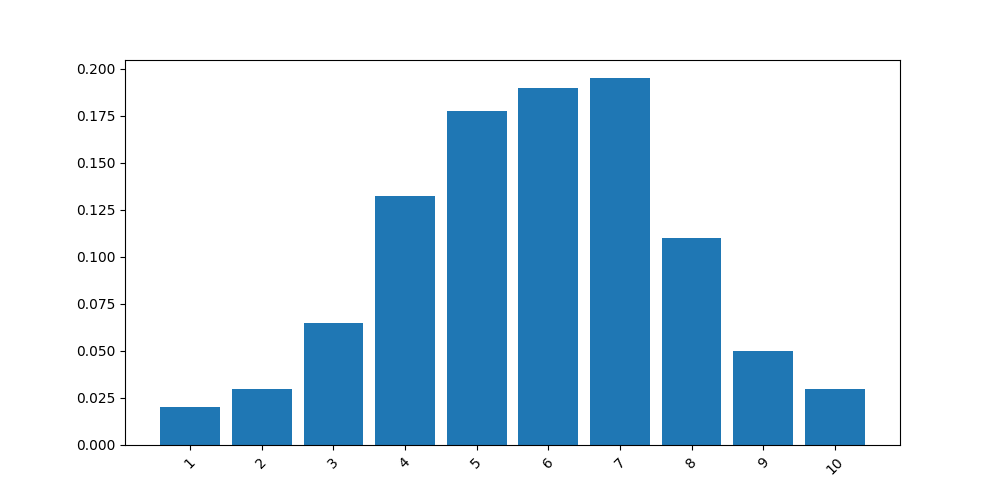}
\\[-0.7em]
prior mood & &
predicted marks
\end{tabular}
\end{center}
\caption{Distributions occurring in Example~\ref{PredictionEx}.}
\label{MoodMarkChannelFig}
\end{figure}

\begin{example}
\label{PredictionEx}
We consider a very simple situation as instantiation of predictive
coding. Assume we use only three possible options to describe the mood
of a teacher, namely: pessimistic ($p$), neutral ($n$) or optimistic
($o$). We thus have a three-element probability space $X = \{p, n,
o\}$. We assume an \emph{a priori} mood distribution:
\[ \begin{array}{rcl}
\sigma
& = &
\frac{1}{8}\ket{p} + \frac{3}{8}\ket{n} + \frac{1}{2}\ket{o}.
\end{array} \]

\noindent This mood thus tends towards optimism.

Associated with these different moods the teacher has different views
on how pupils perform in a particular test. This performance is
expressed in terms of marks, which can range from $1$ to $10$, where
$10$ is best. The probability space for these marks is written as $Y =
\{1,2,\ldots,10\}$.

The view of the teacher is expressed via a channel $c\colon X \chanto
Y$ with:
\[ \begin{array}{rcl}
c(p)
& = &
\frac{1}{50}\ket{1} + \frac{2}{50}\ket{2} + \frac{10}{50}\ket{3} + 
   \frac{15}{50}\ket{4} + \frac{10}{50}\ket{5} +
   \frac{6}{50}\ket{6} + \frac{3}{50}\ket{7} + \frac{1}{50}\ket{8} + 
   \frac{1}{50}\ket{9} + \frac{1}{50}\ket{10}
\\[+0.2em]
c(n)
& = &
\frac{1}{50}\ket{1} + \frac{2}{50}\ket{2} + \frac{4}{50}\ket{3} + 
   \frac{10}{50}\ket{4} + \frac{15}{50}\ket{5} +
   \frac{10}{50}\ket{6} + \frac{5}{50}\ket{7} + \frac{1}{50}\ket{8} + 
   \frac{1}{50}\ket{9} + \frac{1}{50}\ket{10}
\\[+0.2em]
c(o)
& = &
\frac{1}{50}\ket{1} + \frac{1}{50}\ket{2} + \frac{1}{50}\ket{3} + 
   \frac{2}{50}\ket{4} + \frac{4}{50}\ket{5} +
   \frac{10}{50}\ket{6} + \frac{15}{50}\ket{7} + \frac{10}{50}\ket{8} + 
   \frac{4}{50}\ket{9} + \frac{2}{50}\ket{10}.
\end{array} \]


\noindent These three outcomes are plotted in the first row in
Figure~\ref{MoodMarkChannelFig}. They clearly show that the better
the mood, the better the marks.

The second row in Figure~\ref{MoodMarkChannelFig} describes the
mood distribution $\sigma\in\Dst(X)$ and the predicted marks
$c \push \sigma\in\Dst(Y)$, for this mood distribution. The latter
is a convex combination of the three plots in the top row, where
the weights are determined by $\sigma$.

This example will be continued below. The teacher will be confronted
with the marks that the pupils actually obtain. This will lead the
teacher to an update of his/her own mood, in two possible (different)
ways, according to Pearl and according to Jeffrey.
\end{example}

\section{Pearl's updating, increasing what's right}\label{PearlSec}

Before describing Pearl's updating, we collect the relevant notions
and definitions, especially about predicates, validity, updating and
predicate transformations. The new material starts in
Subsection~\ref{PearlSubsec}.

\subsection{Predicates and updating}\label{PredSubsec}

Probabilistic revision involves updating a distribution on the basis
of evidence. Traditionally this evidence takes the form of an event,
that is a subset $E\subseteq X$ of the probability space $X$. A
useful, more general approach uses (fuzzy) predicates as
evidence. They are functions of the form $p\colon X \rightarrow
[0,1]$, with characteristic functions of subsets as a special `sharp'
case. A point predicate, for $x\in X$, is a special (sharp) predicate
$\indic{x} \colon X \rightarrow [0,1]$ sending $x'\neq x$ to $0$ and
$x$ to $1$. Every predicate $p$ on a finite set $X$ can be described
as finite sum $\sum_{x} p(x)\cdot\indic{x}$. We shall write $\Pred(X)
= [0,1]^{X}$ for the set of predicates on $X$.

For a distribution $\omega\in\Dst(X)$ and a predicate $p\in\Pred(X)$
on the same set we write the \emph{validity} of the predicate/evidence
$p$ in the state $\omega$ as $\omega\models p$. It is a number in
$[0,1]$, defined as expected value:
\[ \begin{array}{rcl}
\omega\models p
& \;=\; &
\displaystyle\sum_{x\in X}\, \omega(x)\cdot p(x).
\end{array} \]

\noindent When this validity is non-zero, we define the updated state
$\omega|_{p}\in\Dst(X)$ as the normalised product:
\[ \begin{array}{rcl}
\omega|_{p}(x)
& = &
\displaystyle\frac{\omega(x)\cdot p(x)}{\omega\models p}.
\end{array} \]

\noindent This updating satisfies some important properties, including
Bayes' law and the multiple update law below.
\begin{equation}
\label{BayesEqn}
\begin{array}{rccclcrcl}
\omega|_{p} \models q
& = &
\displaystyle\frac{\omega\models p\andthen q}{\omega\models p}
& = &
\displaystyle\frac{(\omega|_{q}\models p)\cdot (\omega\models q)}
   {\omega\models p}
& \hspace*{8em} &
\omega|_{p}|_{q}
& = &
\omega|_{p\andthen q},
\end{array}
\end{equation}

\noindent where the conjunction $p\andthen q$ is pointwise
multiplication: $(p\andthen q)(x) = p(x) \cdot q(x)$. For more
details,see \textit{e.g.}~\cite{JacobsZ21,Jacobs19c,Jacobs18c}.

Each channel $c \colon X \chanto Y$ gives rise to a predicate
transformation function $\Pred(Y) \rightarrow \Pred(X)$, acting in the
opposite direction. For a predicate $q\colon Y \rightarrow [0,1]$ one
gets $c \pull q \colon X \rightarrow [0,1]$ via:
\[ \begin{array}{rcl}
\big(c \pull q\big)(x)
& = &
\displaystyle\sum_{y\in Y}\, c(x)(y)\cdot q(y).
\end{array} \]

\noindent When a channel is identified with a conditional probability
table in a Bayesian network, as is done by~\cite{JacobsZ21}, this
predicate transformation corresponds to propagation of evidence along
the channel, as path, see~\cite[\S4.3.1]{Pearl88}.

State and predicate transformation $\push$ and $\pull$ are closely related
via validity $\models$, since:
\begin{equation}
\label{ValidityEqn}
\begin{array}{rcl}
(c \push \omega) \models q
& \;=\; &
\omega \models (c \pull q).
\end{array}
\end{equation}

\subsection{Pearl's updating, increasing validity}\label{PearlSubsec}

The idea of an update $\omega|_{p}$ is that the evidence $p$ is
incorporated into the state $\omega$. Hence it is to be expected that
$p$ is `more true' in $\omega|_{p}$ than in $\omega$. That is the
content of the next `update rightness' result, mentioned also
in~\cite{Jacobs19b}, but with a different proof.

\begin{theorem}[Update rightness]
\label{UpdateThm}
For a distribution $\omega$ and a predicate $p$ on the same set,
if the validity $\omega\models p$ is non-zero, one has:
\[ \begin{array}{rcl}
\omega|_{p} \models p
& \;\geq\; &
\omega\models p.
\end{array} \]
\end{theorem}

\begin{myproof}
We show that the difference is non-negative:
\[ \begin{array}[b]{rcl}
\big(\omega|_{p}\models p\big) - \big(\omega\models p\big)
& \smash{\stackrel{\eqref{BayesEqn}}{=}} &
\frac{1}{\omega\models p}\cdot\Big(\omega\models p\andthen p
   \;-\; (\omega\models p)^{2}\Big)
\\[+0.5em]
& = &
\frac{1}{\omega\models p}\cdot\Big(\big(\sum_{x}\omega(x)\cdot p(x)^{2}\big)
   \;-\; 2(\omega\models p)\cdot (\omega\models p) 
   \;+\; (\omega\models p)^{2}\Big)
\\[+0.5em]
& = &
\frac{1}{\omega\models p}\cdot\Big(\big(\sum_{x}\omega(x)\cdot p(x)^{2}\big)
   \;-\; 2\big(\sum_{x}\omega(x)\cdot p(x)\big)\cdot(\omega\models p) 
\\
& & \qquad\qquad \;+\; 
   \big(\sum_{x} \omega(x)\cdot(\omega\models p)^{2}\big)\Big)
\\[+0.5em]
& = &
\frac{1}{\omega\models p}\cdot \sum_{x}\omega(x)\cdot \Big(p(x)^{2}
   \;-\; 2 p(x) \cdot (\omega\models p) 
   \;+\; (\omega\models p)^{2}\Big)
\\[+0.5em]
& = &
\frac{1}{\omega\models p}\cdot\sum_{x}\omega(x)\cdot \Big(p(x) -
   (\omega\models p)\Big)^{2}
\hspace*{\arraycolsep}\geq\hspace*{\arraycolsep}
0.
\end{array} \eqno{\QEDbox} \]
\end{myproof}

We now describe Pearl's update rule, following~\cite{Jacobs19c}.  The
setting is like in predictive coding, as sketched in the introduction,
with a prediction $c \push \sigma$ and a confrontation with external
information. In Pearl's setting this information is evidence, in the
form of a predicate.

\begin{theorem}[Pearl's update]
\label{PearlRuleThm}
Let $c\colon X \chanto Y$ be a channel with a (prior) state
$\sigma\in\Dst(X)$ on its domain $X$. For a predicate $q$ on the
codomain $Y$ of the channel we get an increase of validity
(rightness):
\[ \begin{array}{rclcrcl}
(c \push \sigma_{P}) \models q
& \;\geq\; &
(c \push \sigma) \models q
& \;\mbox{ for the updated/posterior state }\; &
\sigma_{P}
& = &
\sigma|_{c \pull q}.
\end{array} \]

\noindent The update mechanism $\sigma \mapsto \sigma_{P} = \sigma|_{c
  \pull q}$ is Pearl's update rule.
\end{theorem}

\begin{myproof}
By combining Theorem~\ref{UpdateThm} with~\eqref{ValidityEqn} we get:
\[ \hspace*{-0.5em}\begin{array}{rcccccl}
\Big(c \push \big(\sigma|_{c \pull q}\big)\Big) \models q
& \;=\; &
\sigma|_{c \pull q} \models (c \pull q)
& \;\geq\; &
\sigma \models (c \pull q)
& \;=\; &
(c \push \sigma) \models q.
\end{array} \eqno{\QEDbox} \]
\end{myproof}

The update rule of Pearl involves an update with a transformed
predicate. It forms the basis of probabilistic reasoning in Bayesian
networks. Indeed, the conditional probability tables of such networks
are channels and reasoning happens by transforming states and
predicates up and down these channels, in combination with updating at
appropriate points. This perspective comes from~\cite{Pearl88} and is
elaborated by~\cite{JacobsZ21} in channel-based form.

\begin{example}
\label{PearlEx}
We continue in the setting of Example~\ref{PredictionEx} and assume
that the pupils have done rather poorly, with no-one scoring above
$5$, as described by the following evidence/predicate $q$ on the set
of grades $Y = \{1,2,\ldots,10\}$.
\[ \begin{array}{rcl}
q
& = &
\frac{1}{10}\cdot\indic{1} + \frac{3}{10}\cdot\indic{2} + 
\frac{3}{10}\cdot\indic{3} + \frac{2}{10}\cdot\indic{4} + 
\frac{1}{10}\cdot\indic{5}.
\end{array} \]

\noindent The validity of this predicate $q$ in the predicted state $c \push
\sigma$ is:
\[ \begin{array}{rcccccl}
c \push \sigma \models q
& \;=\; &
\sigma \models c \pull q
& \;=\; &
\frac{299}{4000}
& = &
0.07475.
\end{array} \]

\noindent The interested reader may wish to check that the
Pearl-update $\sigma_{P} = \sigma|_{c \pull q}$ and the resulting
increased validity of $q$ are:
\[ \begin{array}{rcl}
\sigma_{P}
& = &
\frac{77}{299}\ket{p} + \frac{162}{299}\ket{n} + \frac{60}{299}\ket{o}
\hspace*{\arraycolsep}\approx\hspace*{\arraycolsep}
0.2575\ket{p} + 0.5418\ket{n} + 0.2007\ket{o}
\\[+0.2em]
c \push \sigma_{P} \models q
& = &
\frac{15577}{149500}
\hspace*{\arraycolsep}\approx\hspace*{\arraycolsep}
0.1042.
\end{array} \]

\auxproof{
We have:
\[ \begin{array}{rcl}
\big(c \pull q\big)(p)
& = &
\sum_{y} c(p)(y)\cdot q(y)
\\
& = &
\frac{1}{50}\cdot\frac{1}{10} + \frac{2}{50}\cdot\frac{3}{10} + 
   \frac{10}{50}\cdot\frac{3}{10} + \frac{15}{50}\cdot\frac{2}{10} + 
   \frac{10}{50}\cdot\frac{1}{10}
\\
& = &
\frac{1 + 6 + 30 + 30 + 10}{500}
\\
& = &
\frac{77}{500}
\\
\big(c \pull q\big)(n)
& = &
\sum_{y} c(n)(y)\cdot q(y)
\\
& = &
\frac{1}{50}\cdot\frac{1}{10} + \frac{2}{50}\cdot\frac{3}{10} + 
   \frac{4}{50}\cdot\frac{3}{10} + \frac{10}{50}\cdot\frac{2}{10} + 
   \frac{15}{50}\cdot\frac{1}{10}
\\
& = &
\frac{1 + 6 + 12 + 20 + 15}{500}
\\
& = &
\frac{54}{500}
\\
\big(c \pull q\big)(o)
& = &
\sum_{y} c(o)(y)\cdot q(y)
\\
& = &
\frac{1}{50}\cdot\frac{1}{10} + \frac{1}{50}\cdot\frac{3}{10} + 
   \frac{1}{50}\cdot\frac{3}{10} + \frac{2}{50}\cdot\frac{2}{10} + 
   \frac{4}{50}\cdot\frac{1}{10}
\\
& = &
\frac{1 + 3 + 3 + 4 + 4}{500}
\\
& = &
\frac{15}{500}.
\end{array} \]

\noindent Hence:
\[ \begin{array}{rcccccl}
c \push \sigma \models q
& = &
\sigma \models c \pull q
& = &
\frac{1}{8}\cdot\frac{77}{500} + \frac{3}{8}\cdot\frac{54}{500} +
   \frac{1}{2}\cdot\frac{15}{500}
& = &
\frac{299}{4000}.
\end{array} \]

\noindent But then:
\[ \begin{array}{rcl}
\sigma|_{c \pull q}
& = &
\displaystyle
\frac{\nicefrac{1}{8}\cdot\nicefrac{77}{500}}{\nicefrac{299}{4000}}\ket{p} 
+ \frac{\nicefrac{3}{8}\cdot\nicefrac{54}{500}}{\nicefrac{299}{4000}}\ket{n}  
+ \frac{\nicefrac{1}{2}\cdot\nicefrac{15}{500}}{\nicefrac{299}{4000}}\ket{o} 
\\[+0.5em]
& = &
\frac{77}{299}\ket{p} + \frac{162}{299}\ket{n} + \frac{60}{299}\ket{o}.
\end{array} \]

\noindent So that:
\[ \begin{array}{rcl}
c \push \big(\sigma|_{c \pull q}\big) \models q
& = &
\sigma|_{c \pull q} \models c \pull q
\\
& = &
\frac{77}{299}\cdot\frac{77}{500} + 
    \frac{162}{299}\cdot\frac{54}{500} + \frac{60}{299}\cdot\frac{15}{500}
\\
& = &
\frac{5929 + 8748 + 900}{149500}
\\
& = &
\frac{15577}{149500}
\\
& \approx &
0.1042.
\end{array} \]
}

\noindent We thus see an increase of validity, roughly from $0.07$ to
$0.10$.
\end{example}

\section{Jeffrey's updating, decreasing wrongness}\label{JeffreySec}

Before we can describe Jeffrey's update rule we need to introduce some
additional background material, about Kullback-Leibler divergence and
about the `dagger' inverse of a channel.

\subsection{Divergence and inversion}\label{DivergenceSubsec}

There are several ways to measure the difference between two
distributions on the same set. Here we shall use the so-called
Kullback-Leibler divergence, which is quite standard. For
$\omega,\rho\in\Dst(X)$ it is defined as:
\[ \begin{array}{rcl}
\KLD\big(\omega,\rho\big)
& = &
\displaystyle\sum_{x\in X}\,\displaystyle \omega(x)\cdot
   \ln\left(\frac{\omega(x)}{\rho(x)}\right).
\end{array} \]

\noindent Here we use the natural logarithm $\ln$, where sometimes the
2-logarithm is used.

One can show that $\KLD(\omega,\rho) \geq 0$ and $\KLD(\omega,\rho) =
0$ if and only if $\omega=\rho$. In general one has $\KLD(\omega,\rho)
\neq \KLD(\rho,\omega)$, so that $\KLD$ is not a metric distance
function. Indeed, it is called divergence and not distance. One can
show that state transformation is divergence-decreasing, in the sense
that $\KLD(c \push \omega, c\push \rho) \leq \KLD(\omega,\rho)$, but we
don't need that property here.

We now look at inversion of a channel $c\colon X \chanto Y$, where we
assume a distribution $\sigma\in\Dst(X)$ on its domain. We turn it
into a channel $Y\chanto X$ in the other direction, written as
$c^{\dag}_{\sigma} \colon Y \chanto X$, and defined as:
\begin{equation}
\label{DaggerEqn}
\begin{array}{rcccccl}
c_{\sigma}^{\dag}(y)(x)
& = &
\sigma|_{c \pull \indic{y}}(x)
& = &
\displaystyle\frac{\sigma(x) \cdot (c \pull \indic{y})(x)}
   {\sigma\models c \pull \indic{y}}
& = &
\displaystyle\frac{\sigma(x) \cdot c(x)(y)}
   {(c \push \sigma)(y)}.
\end{array}
\end{equation}

\noindent The latter formulation shows that we need to require that
the predicted state $c \push \sigma$ has full support.

If channel $c$ represents a conditional probability $p(y\mid x)$, then
its inversion $c^{\dag}_{\sigma}$ corresponds to the Bayesian
inversion $p(x\mid y)$. Such Bayesian inversions play a basic role in
predictive coding, see \textit{e.g.}~\cite{FristonK09}. The dagger
notation is used because this inversion behaves like in so-called
dagger categories, see \cite{ClercDDG17,ChoJ19,Fritz20} for more
information.  Such daggers/inversions are also used to capture the
reversibility of quantum computations via conjugate transposes, see
for further information \textit{e.g.}~\cite{AbramskyC09}.




\section{Jeffrey's updating, decreasing divergence}\label{JeffreyDecreaseSec}

The main result (Theorem~\ref{JeffreyRuleThm}) below states that
Jeffrey's update rule decreases divergence. The proof is non-trivial
and can be found in the appendix.

Recall that in Pearl's updating in Theorem~\ref{PearlRuleThm} the
prediction $c \push \sigma$ is confronted with external evidence, in the
form of a predicate.  In Jeffrey's case one uses an external state
(distribution) as evidence, instead of a predicate.

\begin{theorem}
\label{JeffreyRuleThm}
Let $c\colon X\chanto Y$ be a channel, whose codomain $Y$ is a finite
set, with a state $\sigma\in\Dst(X)$ on its domain, such that the
predicted state $c \push \sigma$ on $Y$ has full support. For an
`evidence' state $\tau\in\Dst(Y)$ there is a reduction of divergence:
\[ \begin{array}{rclcrcl}
\KLD\big(\tau, c \push \sigma_{J}\big)
& \leq &
\KLD\big(\tau, c \push \sigma\big)
& \qquad\mbox{for}\qquad &
\sigma_{J}
& = &
c_{\sigma}^{\dag} \push \tau.
\end{array} \]

\noindent The update mechanism $\sigma \mapsto \sigma_{J} =
c_{\sigma}^{\dag} \push \tau$ is Jeffrey's update rule,
see~\cite{Jacobs19c}. \QED
\end{theorem}

There is an earlier result describing the effect of Jeffrey's update
rule, given for instance by~\cite[Prop.~3.11.2]{Halpern03}. Translated
to the current context it describes the divergence between the orginal
state $\sigma$ and its Jeffrey update as infimum:
\[ \begin{array}{rcl}
\KLD\big(\sigma, \klin{f}_{\sigma}^{\dag} \push \tau\big)
& = &
\displaystyle\bigwedge\set{\KLD\big(\sigma, \omega\big)}{
   \omega\in\Dst(X) \mbox{ with } \klin{f} \push \omega = \tau}.
\end{array} \]

\noindent Recall that $\klin{f} \colon X \chanto Y$ is the promotion
of an ordinary function $f\colon X \rightarrow Y$ to a deterministic
channel, with $\klin{f}(x) = 1\ket{f(x)}$. Indeed, this earlier result
is restricted, since it only works for deterministic channels, and not
for channels in general, like Theorem~\ref{JeffreyRuleThm}.  In the
deterministic case things are easy: one can update $\sigma$ to $\sigma' =
\klin{f}_{\sigma}^{\dag} \push \tau$ and get a perfect prediction,
since $\klin{f} \push \sigma' = \tau$.

\auxproof{
We first show $\klin{f} \push \big(\klin{f}_{\sigma}^{\dag} \push
\tau\big) = \tau$. This shows $(\geq)$.
\[ \begin{array}{rcl}
\Big(\klin{f} \push \big(\klin{f}_{\sigma}^{\dag} \push \tau\big)\Big)(y)
& = &
\displaystyle\sum_{x\in f^{-1}(y)} \big(\klin{f}_{\sigma}^{\dag} \push \tau\big)(x)
\\[+1em]
& = &
\displaystyle\sum_{x\in f^{-1}(y)} \, \sum_{z\in Y} \,
   \tau(z) \cdot \klin{f}_{\sigma}^{\dag}(z)(x)
\\[+1em]
& = &
\displaystyle\sum_{x\in f^{-1}(y)} \, \sum_{z\in Y} \,
   \tau(z) \cdot \sigma|_{\klin{f} \pull \indic{z}}(x)
\\[+1em]
& = &
\displaystyle\sum_{x\in f^{-1}(y)} \, \sum_{z\in Y} \,
   \tau(z) \cdot \frac{\sigma(x) \cdot (\klin{f} \pull \indic{z})(x)}
   {\sigma \models \klin{f} \pull \indic{z}}
\\[+1em]
& = &
\displaystyle\sum_{x\in f^{-1}(y)} \, \sum_{z\in Y} \,
   \tau(z) \cdot \frac{\sigma(x) \cdot \indic{z}(f(x))}
   {\klin{f} \push \sigma \models \indic{z}}
\\[+1em]
& = &
\displaystyle\sum_{x\in f^{-1}(y)} \, 
   \tau(y) \cdot \frac{\sigma(x)}{\Dst(f)(\sigma)(y)}
\\[+1.3em]
& = &
\displaystyle \tau(y) \cdot \frac{\Dst(f)(\sigma)(y)}{\Dst(f)(\sigma)(y)}
\\[+0.5em]
& = &
\tau(y).
\end{array} \]

Let $\klin{f} \push \omega = \tau$. Then:
\[ \begin{array}{rcl}
\lefteqn{\KLD\big(\sigma, \klin{f}_{\sigma}^{\dag} \push \tau\big) \;-\;
   \KLD\big(\sigma, \omega\big)}
\\
& = &
\displaystyle\sum_{x\in X}\, \sigma(x) \cdot \left[
   \ln\left(\frac{\sigma(x)}{(\klin{f}_{\sigma}^{\dag} \push \tau)(x)}\right) \;-\;
   \ln\left(\frac{\sigma(x)}{\omega(x)}\right)\right]
\\[+1em]
& = &
\displaystyle\sum_{x\in X}\, \sigma(x) \cdot 
   \ln\left(\frac{\sigma(x)}{(\klin{f}_{\sigma}^{\dag} \push \tau)(x)} \cdot
   \frac{\omega(x)}{\sigma(x)}\right)
\\[+1em]
& \leq &
\displaystyle \ln\left(\sum_{x\in X}\, 
   \frac{\sigma(x)\cdot\omega(x)}{(\klin{f}_{\sigma}^{\dag} \push \tau)(x)}\right)
   \quad\mbox{by Jensen's inequality}
\\[+1em]
& \leq &
\ln(1) \quad \mbox{see below}
\\
& = &
0.
\end{array} \]

\noindent The remaining step:
\[ \begin{array}{rcl}
\displaystyle\sum_{x\in X}\, \frac{\sigma(x)\cdot\omega(x)}
   {(\klin{f}_{\sigma}^{\dag} \push \tau)(x)}
& = &
\displaystyle\sum_{x\in X}\, \frac{\sigma(x)\cdot\omega(x)}
   {\sum_{y} \tau(y) \cdot \sigma|_{\klin{f} \pull \indic{y}}(x)}
\\[+1em]
& = &
\displaystyle\sum_{x\in X}\, \frac{\sigma(x)\cdot\omega(x)}
   {\sum_{y} \tau(y) \cdot \frac{\sigma(x) \cdot (\klin{f} \pull \indic{y})(x)}
   {\sigma \models \klin{f} \pull \indic{y}}}
\\[+1em]
& = &
\displaystyle\sum_{x\in X}\, \frac{\omega(x)}
   {\sum_{y} \tau(y) \cdot \frac{\indic{y}(f(x))}{\Dst(f)(\sigma)(y)}}
\\[+1em]
& = &
\displaystyle\sum_{x\in X}\, \frac{\omega(x)}
   {\frac{\tau(f(x))}{\Dst(f)(\sigma)(f(x))}}
\\[+1em]
& = &
\displaystyle\sum_{x\in X}\, \frac{\omega(x) \cdot \Dst(f)(\sigma)(f(x))}
   {\tau(f(x))}
\\[+1em]
& = &
\displaystyle\sum_{x\in X}\, \frac{\omega(x) \cdot \Dst(f)(\sigma)(f(x))}
   {(\klin{f} \push \omega)(f(x))}
\\[+1em]
& = &
\displaystyle\sum_{x\in X}\, \frac{\omega(x)}
   {\sum_{x'\in f^{-1}(f(x))} \omega(x')} \cdot \Dst(f)(\sigma)(f(x))
\\[+1em]
& \leq &
\displaystyle\sum_{x\in X}\, \Dst(f)(\sigma)(f(x))
\\[+1em]
& \leq &
\displaystyle\sum_{y\in Y}\, \Dst(f)(\sigma)(y)
\\
& = &
1.
\end{array} \]
}

Further, we mention that the update rules of Pearl and Jeffrey
are interdefinable, see~\cite{ChanD05} or~\cite{Jacobs19c}.
Let $c\colon X \chanto Y$ and $\sigma\in\Dst(X)$ be given. For an
evidence predicate $q$ we can obtain Pearl's update on the left
below, in terms of Jeffrey's update on the right:
\[ \begin{array}{rcl}
\sigma|_{c \pull q}
& = &
c_{\sigma}^{\dag} \push \Big((c \push \sigma)\big|_{q}\Big).
\end{array} \]

\noindent This uses the predicted state $c \push \sigma$, updated with
the Pearl-evidence $q$, as Jeffrey-evidence. Similarly, if we have
an evidence state $\tau\in\Dst(Y)$ we have:
\[ \begin{array}{rclcrcl}
c_{\sigma}^{\dag} \push \tau
& = &
\sigma|_{c \pull q} 
& \quad\mbox{for}\quad &
q(y)
& = &
\displaystyle\frac{\tau(y)}{(c\push \sigma)(y)}.
\end{array} \]

\noindent One may need to rescale the fraction to obtain a predicate, but
such rescaling does not affect updating.

We take another look at our leading example, but now from Jeffrey's
perspective.

\begin{figure}
\begin{center}
\begin{tabular}{ccccc}
\includegraphics[width=4cm]{mood}
& \qquad\quad & 
\includegraphics[width=4cm]{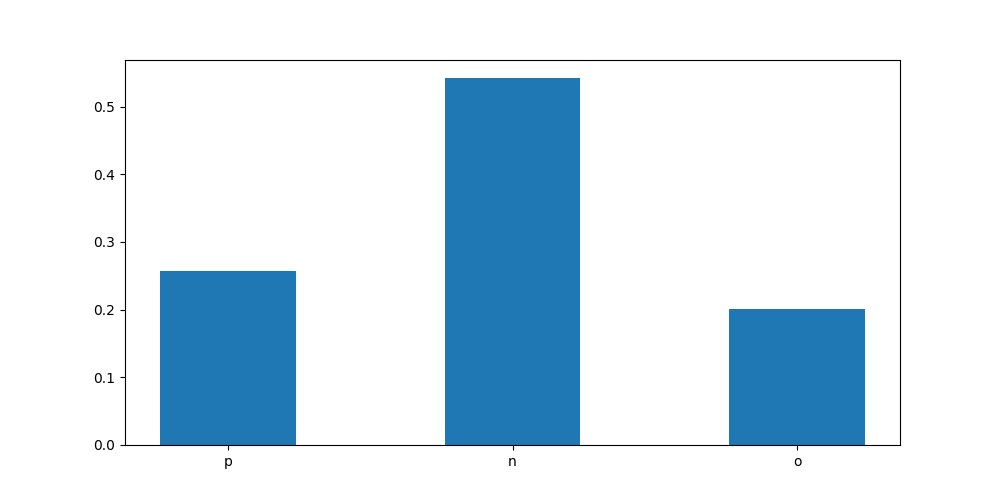}
& \qquad\quad &
\includegraphics[width=4cm]{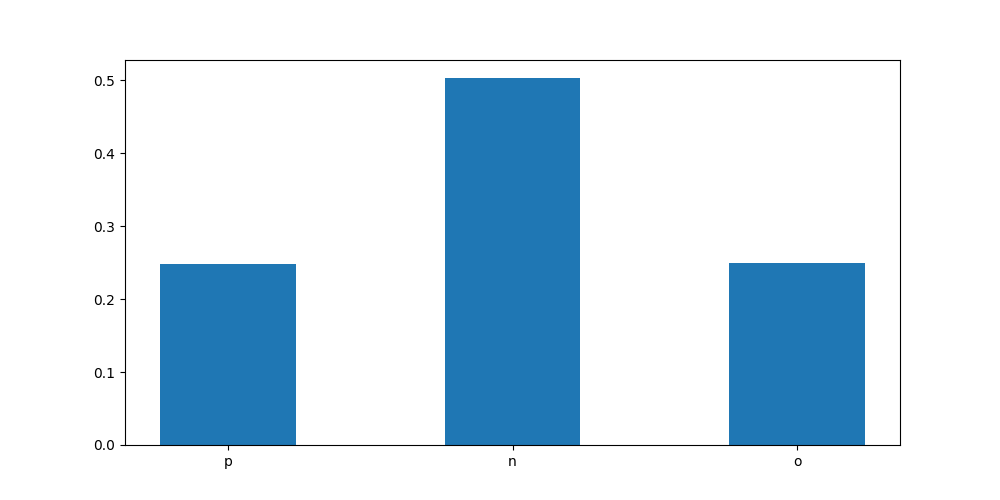}
\\[-0.7em]
prior mood & & 
mood after Pearl-update & & 
mood after Jeffrey-update
\\[-0.3em]
$\scriptstyle 0.125\ket{p} + 0.375\ket{n} + 0.5\ket{o}$
& &
$\scriptstyle 0.2575\ket{p} + 0.5418\ket{n} + 0.2007\ket{o}$
& &
$\scriptstyle 0.2486\ket{p} + 0.5025\ket{n} + 0.2489\ket{o}$
\end{tabular}
\end{center}
\caption{Mood updates from Examples~\ref{PredictionEx} and~\ref{JeffreyEx}.}
\label{MoodUpdateFig}
\end{figure}

\begin{example}
\label{JeffreyEx}
Recall the situation of Example~\ref{PredictionEx}, with a teacher
predicting the performance of pupils, depending on the teacher's mood.
The evidence predicate $q$ from Example~\ref{PearlEx} can be
translated into a state $\tau$ on the set $G$ of grades:
\[ \begin{array}{rcl}
\tau
& = &
\frac{1}{10}\ket{1} + \frac{3}{10}\ket{2} + \frac{3}{10}\ket{3} + 
   \frac{2}{10}\ket{4} + \frac{1}{10}\ket{5}.
\end{array} \]

\noindent There is an a priori divergence $\KLD(\tau, c \push \sigma)
\approx 1.336$. With some effort one can prove that the Jeffrey-update
of $\sigma$ is:
\[ \begin{array}{rcccl}
\sigma_{J}
\hspace*{\arraycolsep}=\hspace*{\arraycolsep}
c_{\sigma}^{\dag} \push \tau
& = &
\frac{972795}{3913520}\ket{p} + \frac{1966737}{3913520}\ket{n} +
    \frac{973988}{3913520}\ket{o}
& \approx &
0.2486\ket{p} + 0.5025\ket{n} + 0.2489\ket{o}.
\end{array} \]

\noindent The divergence has now dropped, from $1.336$ to $\KLD(\tau,
c\push\sigma_{J}) \approx 1.087$.

\auxproof{
In order to compute the updated state $\sigma_{J} = c_{\sigma}^{\dag}
\push \tau$ we compute consecutively:
\[ \begin{array}{rcl}
\sigma \models c \pull \indic{1}
& = &
\sum_{x} \sigma(x) \cdot c(x)(1)
\hspace*{\arraycolsep}=\hspace*{\arraycolsep}
\frac{1}{8}\cdot\frac{1}{50} + \frac{3}{8}\cdot\frac{1}{50} + 
   \frac{1}{2}\cdot\frac{1}{50}
\hspace*{\arraycolsep}=\hspace*{\arraycolsep}
\frac{1}{50}
\\
\sigma|_{c \pull \indic{1}}
& = &
\displaystyle
\frac{\nicefrac{1}{8}\cdot\nicefrac{1}{50}}{\nicefrac{1}{50}}\ket{p} 
+ \frac{\nicefrac{3}{8}\cdot\nicefrac{1}{50}}{\nicefrac{1}{50}}\ket{n}  
+ \frac{\nicefrac{1}{2}\cdot\nicefrac{1}{50}}{\nicefrac{1}{50}}\ket{o} 
\hspace*{\arraycolsep}=\hspace*{\arraycolsep}\textstyle
\frac{1}{8}\ket{p} + \frac{3}{8}\ket{n} + \frac{1}{2}\ket{o}
\\
\sigma \models c \pull \indic{2}
& = &
\frac{1}{8}\cdot\frac{2}{50} + \frac{3}{8}\cdot\frac{2}{50} + 
   \frac{1}{2}\cdot\frac{1}{50}
\hspace*{\arraycolsep}=\hspace*{\arraycolsep}
\frac{3}{100}
\\
\sigma|_{c \pull \indic{2}}
& = &
\displaystyle
\frac{\nicefrac{1}{8}\cdot\nicefrac{2}{50}}{\nicefrac{3}{100}}\ket{p} 
+ \frac{\nicefrac{3}{8}\cdot\nicefrac{2}{50}}{\nicefrac{3}{100}}\ket{n}  
+ \frac{\nicefrac{1}{2}\cdot\nicefrac{1}{50}}{\nicefrac{3}{100}}\ket{o} 
\hspace*{\arraycolsep}=\hspace*{\arraycolsep}\textstyle
\frac{1}{6}\ket{p} + \frac{1}{2}\ket{n} + \frac{1}{3}\ket{o}
\\
\sigma \models c \pull \indic{3}
& = &
\frac{1}{8}\cdot\frac{10}{50} + \frac{3}{8}\cdot\frac{4}{50} + 
   \frac{1}{2}\cdot\frac{1}{50}
\hspace*{\arraycolsep}=\hspace*{\arraycolsep}
\frac{13}{200}
\\
\sigma|_{c \pull \indic{3}}
& = &
\displaystyle
\frac{\nicefrac{1}{8}\cdot\nicefrac{10}{50}}{\nicefrac{13}{200}}\ket{p} 
+ \frac{\nicefrac{3}{8}\cdot\nicefrac{4}{50}}{\nicefrac{13}{200}}\ket{n}  
+ \frac{\nicefrac{1}{2}\cdot\nicefrac{1}{50}}{\nicefrac{13}{200}}\ket{o} 
\hspace*{\arraycolsep}=\hspace*{\arraycolsep}\textstyle
\frac{5}{13}\ket{p} + \frac{6}{13}\ket{n} + \frac{2}{13}\ket{o}
\\
\sigma \models c \pull \indic{4}
& = &
\frac{1}{8}\cdot\frac{15}{50} + \frac{3}{8}\cdot\frac{10}{50} + 
   \frac{1}{2}\cdot\frac{2}{50}
\hspace*{\arraycolsep}=\hspace*{\arraycolsep}
\frac{53}{400}
\\
\sigma|_{c \pull \indic{4}}
& = &
\displaystyle
\frac{\nicefrac{1}{8}\cdot\nicefrac{15}{50}}{\nicefrac{53}{400}}\ket{p} 
+ \frac{\nicefrac{3}{8}\cdot\nicefrac{4}{50}}{\nicefrac{53}{400}}\ket{n}  
+ \frac{\nicefrac{1}{2}\cdot\nicefrac{2}{50}}{\nicefrac{53}{400}}\ket{o} 
\hspace*{\arraycolsep}=\hspace*{\arraycolsep}\textstyle
\frac{15}{53}\ket{p} + \frac{30}{53}\ket{n} + \frac{8}{53}\ket{o}
\\
\sigma \models c \pull \indic{5}
& = &
\frac{1}{8}\cdot\frac{10}{50} + \frac{3}{8}\cdot\frac{15}{50} + 
   \frac{1}{2}\cdot\frac{4}{50}
\hspace*{\arraycolsep}=\hspace*{\arraycolsep}
\frac{71}{400}
\\
\sigma|_{c \pull \indic{5}}
& = &
\displaystyle
\frac{\nicefrac{1}{8}\cdot\nicefrac{10}{50}}{\nicefrac{71}{400}}\ket{p} 
+ \frac{\nicefrac{3}{8}\cdot\nicefrac{15}{50}}{\nicefrac{71}{400}}\ket{n}  
+ \frac{\nicefrac{1}{2}\cdot\nicefrac{4}{50}}{\nicefrac{71}{400}}\ket{o} 
\hspace*{\arraycolsep}=\hspace*{\arraycolsep}\textstyle
\frac{10}{71}\ket{p} + \frac{45}{71}\ket{n} + \frac{16}{71}\ket{o}
\end{array} \]

\noindent We can now compute:
\[ \begin{array}{rcl}
\sigma_{J}
& = &
c_{\sigma}^{\dag} \push \tau
\\
& = &
\frac{1}{10}\Big(\frac{1}{8}\ket{p} + \frac{3}{8}\ket{n} + \frac{1}{2}\ket{o}\Big)
+ \frac{3}{10}\Big(\frac{1}{6}\ket{p} + \frac{1}{2}\ket{n} + \frac{1}{3}\ket{o}\Big)
\\
& & \qquad +\; 
\frac{3}{10}\Big(\frac{5}{13}\ket{p} + \frac{6}{13}\ket{n} + \frac{2}{13}\ket{o}\Big)
+ \frac{2}{10}\Big(\frac{15}{53}\ket{p} + \frac{30}{53}\ket{n} + \frac{8}{53}\ket{o}\Big)
\\
& & \qquad +\; 
\frac{1}{10}\Big(\frac{10}{71}\ket{p} + \frac{45}{71}\ket{n} + \frac{16}{71}\ket{o}\Big)
\\
& = &
\Big(\frac{1}{10}\cdot\frac{1}{8} + \frac{3}{10}\cdot\frac{1}{6} +
   \frac{3}{10}\cdot\frac{5}{13} + \frac{2}{10}\cdot\frac{15}{53} +
   \frac{1}{10}\cdot\frac{10}{71}\Big)\ket{p}
\\
& & \qquad +\; 
\Big(\frac{1}{10}\cdot\frac{3}{8} + \frac{3}{10}\cdot\frac{1}{2} +
   \frac{3}{10}\cdot\frac{6}{13} + \frac{2}{10}\cdot\frac{30}{53} +
   \frac{1}{10}\cdot\frac{45}{71}\Big)\ket{n}
\\
& & \qquad +\; 
\Big(\frac{1}{10}\cdot\frac{1}{2} + \frac{3}{10}\cdot\frac{1}{3} +
   \frac{3}{10}\cdot\frac{2}{13} + \frac{2}{10}\cdot\frac{8}{53} +
   \frac{1}{10}\cdot\frac{16}{71}\Big)\ket{n}
\\
& = &
\frac{972795}{3913520}\ket{p} + \frac{1966737}{3913520}\ket{n} +
    \frac{973988}{3913520}\ket{o}.
\end{array} \]
}

In the end it is interesting to compare the original (prior) mood with
its Pearl- and Jeffrey-updates. In Figure~\ref{MoodUpdateFig} the
prior mood is reproduced from Figure~\ref{MoodMarkChannelFig}, for
easy comparison. The Pearl and Jeffrey updates differ only slightly,
to be precise with $\KLD(\sigma_{P}, \sigma_{J}) \approx 0.007$. They
both show that the bad grades evidence deteriorates the teacher's
mood. Examples where the Pearl- and Jeffrey-updates differ wildly are
given by~\cite{Jacobs19c}.
\end{example}

\begin{remark}
\label{ExclusivityRem}
At this stage one could say: well, fair enough, so Pearl's rule
increases validity and Jeffrey's rule decreases divergence, but how
``exclusive'' are these results?  Maybe there is also a decrease of
divergence in Pearl's updating and an increase of validity in
Jeffrey's updating.

Recall that Pearl's rule uses a predicate $q$ as evidence and
Jeffrey's rules uses a state $\tau$. If we keep states and predicates
apart, as mathematical entities of different types, there is no way to
express a divergence-decrease in Pearl's setting or a
validity-increase in Jeffrey's setting.

Nevertheless, in Examples~\ref{PearlEx} and~\ref{JeffreyEx} we have
seen that the evidence $q$ and $\tau$ are basically the same. In such
a situation we can show that Pearl's update rule need not give a
divergence-decrease and Jeffrey's rule need not produce a
validity-increase. We give an example which demonstrates both points
at the same time.

Take sets $X = \{0,1\}$ and $Y = \{a,b,c\}$ with uniform prior $\sigma
= \frac{1}{2}\ket{0} + \frac{1}{2}\ket{1} \in \Dst(X)$. We use the
channel $c\colon X \chanto Y$ given by:
\[ \begin{array}{rclcrcl}
c(0)
& = &
\frac{1}{9}\ket{a} + \frac{2}{3}\ket{b} + \frac{2}{9}\ket{c}
& \qquad\mbox{and}\qquad &
c(1)
& = &
\frac{7}{25}\ket{a} + \frac{7}{25}\ket{b} + \frac{11}{25}\ket{c}.
\end{array} \]

\noindent The predicted state is then $c \push \sigma =
\frac{44}{225}\ket{a} + \frac{71}{150}\ket{b} +
\frac{149}{450}\ket{c}$. We use as `equal' evidence predicate and
state:
\[ \begin{array}{rclcrcl}
q
& = &
\frac{1}{2}\cdot\indic{a} + \frac{1}{3}\cdot\indic{b} + 
   \frac{1}{6}\cdot\indic{c}
& \qquad\mbox{and}\qquad &
\tau
& = &
\frac{1}{2}\ket{a} + \frac{1}{3}\ket{b} + \frac{1}{6}\ket{c}.
\end{array} \]

\noindent We then get the following updates, according to Pearl and
Jeffrey, respectively:
\[ \begin{array}{rclcrcl}
\sigma_{P}
& = &
\frac{425}{839}\ket{0} + \frac{414}{839}\ket{1}
& \qquad\mbox{and}\qquad &
\sigma_{J}
& = &
\frac{805675}{1861904}\ket{0} + \frac{1056229}{1861904}\ket{1} 
\\[+0.2em]
& \approx &
0.5066\ket{0} + 0.4934\ket{1}
& & 
& \approx &
0.4327\ket{0} + 0.5673\ket{1}.
\end{array} \]

\noindent The validities and divergences are summarised in the
following tables.
\begin{center}
\begin{tabular}{c||c|c}
\textbf{description} & \textbf{formula} & \textbf{value}
\\
\hline\hline
prior validity & $c \push \sigma \models q$ & $0.31074$
\\
after Pearl & $c \push \sigma_{P} \models q$ & $0.31079$
\\
after Jeffrey & $c \push \sigma_{J} \models q$ & $0.31019$
\end{tabular}
\hspace*{3em}
\begin{tabular}{c||c|c}
\textbf{description} & \textbf{formula} & \textbf{value}
\\
\hline\hline
prior divergence & $\KLD(\tau, c \push \sigma)$ & $0.238$
\\
after Pearl & $\KLD(\tau, c \push \sigma_{P})$ & $0.240$
\\
after Jeffrey & $\KLD(\tau, c \push \sigma_{J})$ & $0.221$
\end{tabular}
\end{center}

\noindent The differences are small, but relevant. Pearl's updating
increases validity, as Theorem~\ref{UpdateThm} dictates, but Jeffrey's
updating does not, in this example. Similarly, Jeffrey's updating
decreases divergence, in line with Theorem~\ref{JeffreyRuleThm}, but
Pearl's updating does not.

One can ask if a divergence-decrease also happens for other forms of
divergence (or distance) between distributions. We do not have an
exhaustive answer but we do know that this fails for total variantion
distance. If we replace the above channel $c$ by $c'$ with $c'(0) =
\frac{1}{10}\ket{a} + \frac{1}{2}\ket{b} + \frac{2}{5}\ket{c}$ and
$c'(1) = \frac{11}{100}\ket{a} + \frac{33}{100}\ket{b} +
\frac{56}{100}\ket{c}$, and keep everything else as it is, then both
Pearl's and Jeffrey's update rule produce an increase of the total
variation distance.
\end{remark}

\section{Application to predictive coding}\label{PredCodSec}

This section first explains how our main result,
Theorem~\ref{JeffreyRuleThm}, can be seen as strengthening the
mathematical basis of predictive coding theory. Then, it illustrates
(for the runing example) how the current framework can be extended
with selective focus and managed expectations.

\subsection{Going beyond free energy for point observations}

The concept of free energy plays an important role in predictive
coding. We briefly describe how it fits into the current framework,
see also \textit{e.g.}~\cite{Bogacz17,Friston09,Friston10}. Free
energy has its basis in statistical physics, going back to Ludwig
Bolzmann in the 19th century, where it is used to describe a thermal
equilibrium in gases. In predictive coding the human mind is also seen
as striving for an equilibrium by reducing prediction errors. This
reduction can be achieved either by internally updating the state or
by externally performing an action. Using the notation of
Theorem~\ref{JeffreyRuleThm}, the former involves changing the
internal state $\sigma$ and the latter involves changing the external
state $\tau$ by taking action. Here we only look at (internal)
updating.

In our set-up in Theorem~\ref{JeffreyRuleThm} we describe an internal
update $\sigma \mapsto \sigma_{J}$ triggered by confrontation with an
external state $\tau$. In the predictive coding framework free energy
is described not with respect to an entire distribution
$\tau\in\Dst(Y)$, but with respect to a single, point observation
$y\in Y$ only. One often thinks of this $y$ as a sample from some
external distribution $\tau$. This single observation $y$ corresponds
to a point state $1\ket{y}$ and the resulting Jeffrey update is:
\begin{equation}
\label{JeffreyPointEqn}
\begin{array}{rcccl}
c_{\sigma}^{\dag} \push 1\ket{y}
& = &
c_{\sigma}^{\dag}(y)
& \smash{\stackrel{\eqref{DaggerEqn}}{=}} &
\displaystyle\sum_{x\in X}\,\displaystyle
   \frac{\sigma(x)\cdot c(x)(y)}{(c \push \sigma)(y)}\bigket{x}.
\end{array}
\end{equation}

\noindent Calculating this distribution may be computationally
demanding, especially in the setting of continuous probability. It is
in particular the normalising factor $(c \push \sigma)(y)$ that one
wishes to avoid. Hence the strategy is not to
compute~\eqref{JeffreyPointEqn} but to find a state $\omega$ that
diverges minimally from~\eqref{JeffreyPointEqn}.  One thus looks for
$\omega$ with minimal divergence:
\[ \begin{array}{rcl}
\KLD\big(\omega, \, c_{\sigma}^{\dag}(y)\big)
& = &
\displaystyle\sum_{x\in X}\, \displaystyle\omega(x)\cdot\ln\left(
   \frac{\omega(x)\cdot (c \push \sigma)(y)}{\sigma(x) \cdot c(x)(y)}\right)
\\[+1em]
& = &
\displaystyle\sum_{x\in X}\, \displaystyle\omega(x)\cdot\ln\left(
   \frac{\omega(x)}{\sigma(x) \cdot c(x)(y)}\right)
   + \sum_{x\in X}\, \omega(x)\cdot
   \ln\Big((c \push \sigma)(y)\Big)
\\[+1em]
& = &
-\mathcal{F}(\omega) + \ln\Big((c \push \sigma)(y)\Big),
\end{array} \]

\noindent where $\mathcal{F}$ is the free energy, defined as:
\[ \begin{array}{rclcrcl}
\mathcal{F}(\omega)
& = &
\displaystyle\sum_{x\in X}\, \displaystyle\omega(x)\cdot\ln\left(
   \frac{\gamma(x,y)}{\omega(x)}\right)
& \qquad\mbox{with joint state}\qquad &
\gamma(x,y)
& = &
\sigma(x)\cdot c(x)(y).
\end{array} \]

\noindent This joint state $\gamma\in\Dst(X\times Y)$ is the
generative model associated with the state and channel $\sigma,
c$. Thus, by finding $\omega$ with maximal free energy
$\mathcal{F}(\omega)$ one obtains at the same time a state $\omega$
that diverges minimally from the Jeffrey update $c^{\dag}_{\sigma} \push
1\ket{y}$. This works since the normalisation factor $(c \push
\sigma)(y)$ does not depend on $\omega$ and can thus by ignored in the
optimalisation.

Our Theorem~\ref{JeffreyRuleThm} enriches predictive coding theory and
shows how to go beyond point observations $y\in Y$, and use a
distribution $\tau\in\Dst(Y)$ as external evidence. Updating with such
$\tau$ as evidence reduces divergence --- and thus prediction error.

\subsection{Incorporating focus and expectation management}

We briefly discuss how the situation in this paper, with a channel
$c\colon X \chanto Y$ mediating between an internal world $X$ and an
external world $Y$ can be used to incorporate aspects of focus and
expectation management (preparation). This is done via appropriately
placed updates, and can be considered both from Pearl's and from
Jeffrey's perspective, each with their own objectives. We illustrate
how this can be done mathematically, without any cognitive claims.

We continue Examples~\ref{PredictionEx}, \ref{PearlEx}
and~\ref{JeffreyEx} with evidence on the set of marks $Y =
\{1,2,\ldots,10\}$, either in the form of a predicate $q$ (Pearl) or a
state $\tau$ (Jeffrey). Suppose the teacher focuses on the bad
grades. This focus may be a conscious decision or instruction, or may
happen unconsciously, \emph{e.g.}~through some form of bias or tunnel
vision. We illustrate how the focus can happen via a subset/event
$F\subseteq Y$, say $F = \{1,2,3\}$ containing bad marks. We write
$\indic{F} \colon Y \rightarrow [0,1]$ for the associated sharp
predicate, with $\indic{F}(y) = 1$ if $y\in F$ and $\indic{F}(y) = 0$
if $y\not\in F$. We show how to use this focus predicate $\indic{F}$
for a tunnel vision on the external world, by incorporating it in the
following manner.
\begin{itemize}
\item Pearl: update internal state $\sigma$ to $\sigma|_{c \,\pull\, (q
  \,\andthen\, \indic{F})}$.

\item Jeffrey: update $\sigma$ to $c_{\sigma}^{\dag} \push (\tau|_{\indic{F}})$
\end{itemize}

\begin{figure}
\begin{center}
\begin{tabular}{ccc}
\includegraphics[width=4cm]{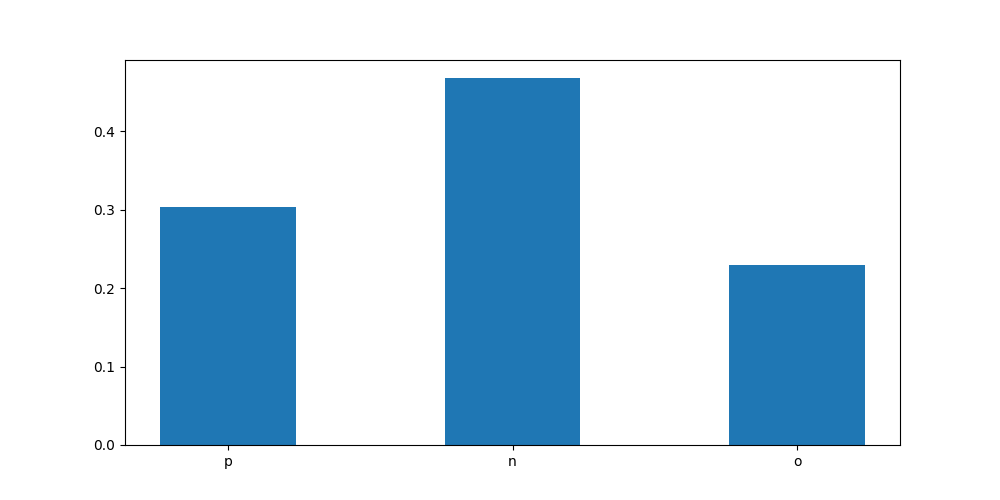}
& \hspace*{5em} &
\includegraphics[width=4cm]{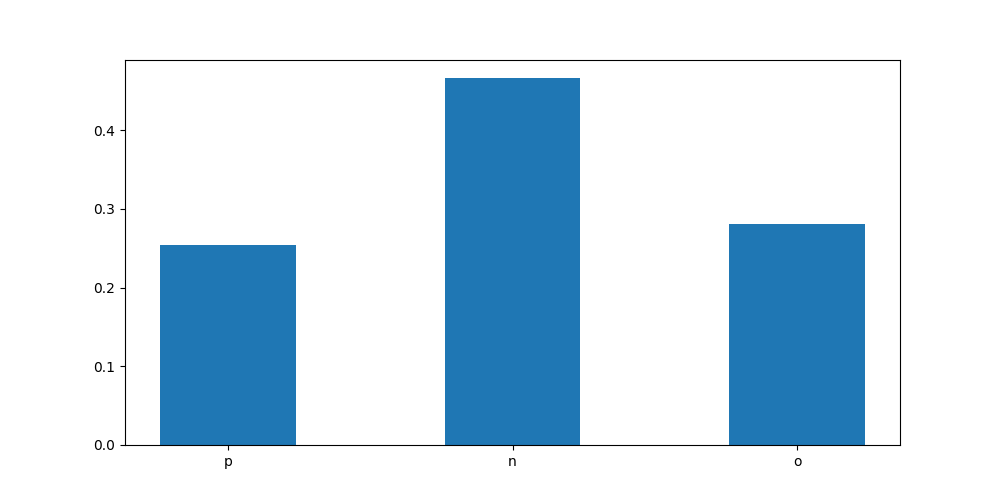}
\\[-0.7em]
Pearl-update & & Jeffrey-update
\\[-0.3em]
$\scriptstyle 0.3033\ket{p} + 0.4672\ket{n} + 0.2295\ket{o}$
& &
$\scriptstyle 0.2541\ket{p} + 0.4657\ket{n} + 0.2802\ket{o}$
\end{tabular}
\end{center}
\caption{Mood updates with focus.}
\label{MoodFocusUpdateFig}
\end{figure}

\noindent In Pearl's case the focus predicate $\indic{F}$ is combined
with the evidence $q$ via conjunction. In Jeffrey's case the focus
predicate is used to update the external state $\tau$ to
$\tau|_{\indic{F}}$ --- which itself may be understood as an action in
predictive coding theory. In both cases the effect is that the
evidence is masked (restricted). The resulting updates of the mood
state $\sigma$ are described in Figure~\ref{MoodFocusUpdateFig}. With
respect to the updates without focus in Figure~\ref{MoodUpdateFig}
there is more pessimism, as a result of the focus on the bad
marks. This pessimistic shift is greater in Pearl's update.

In predictive coding theory, and more generally in cognition theory,
the notion of attention plays an important role, see
\textit{e.g.}~\cite{SchrogerMS15,Clark16}. It refers to the processing
of the prediction error. Above we have deliberately used the informal
term focus, to avoid confusion.

\smallskip

Instead of masking the external evidence via a predicate on the
outside world $Y$ the teacher can also prepare for the outcome by
updating his/her internal state $\sigma$ before adapting to the
external evidence.  This can be seen as a form of managing one's own
expectations. Consider for instance the predicate $r$ on the set $X =
\{p, n, o\}$ of mood options given by:
\[ \begin{array}{rcl}
r
& = &
\frac{7}{10}\cdot\indic{p} + \frac{1}{2}\cdot\indic{n} + 
   \frac{3}{10}\cdot\indic{o}. 
\end{array} \]

\noindent Clearly, it favours pessimism. We can incorporate this
predicate $r$ on the internal side, both in Pearl's and in Jeffrey's
approach:
\begin{itemize}
\item Pearl: update internal state $\sigma$ to $\sigma|_{r}|_{c \pull q}
  = \sigma|_{r\andthen (c \pull q)}$.

\item Jeffrey: update $\sigma$ to $c_{\sigma|_{r}}^{\dag} \push \tau$.
\end{itemize}

\begin{figure}
\begin{center}
\begin{tabular}{c||ccc}
\begin{tabular}{c} \textbf{ bad marks } \\[-0.4em]
   \textbf{ mood update } \end{tabular} & \textbf{Pearl} & & \textbf{Jeffrey}
\\
\hline\hline
\begin{tabular}{c} \textbf{ after } \\[-0.4em]
   \textbf{ pessimistic } \\[-0.4em]
   \textbf{ preparation } \end{tabular} &
$\vcenter{\hbox{\includegraphics[width=4cm]{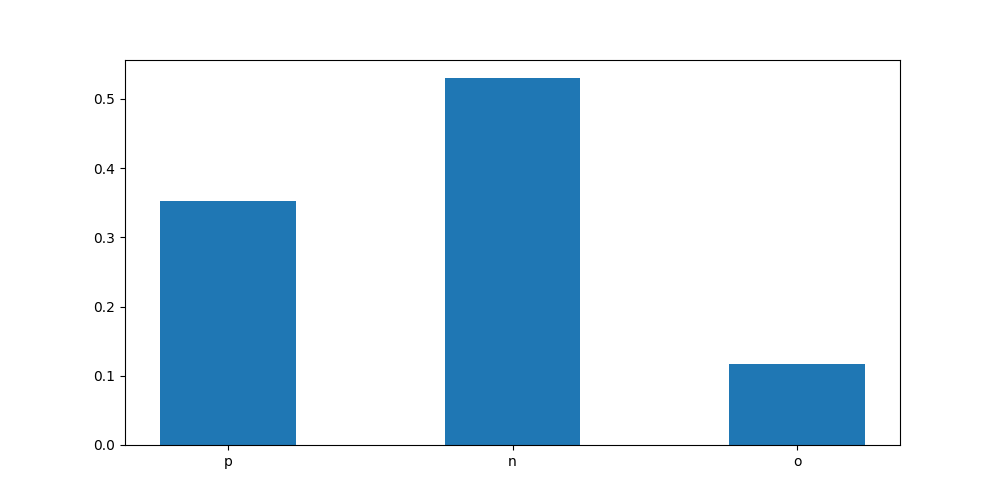}}}$
& \hspace*{1em} &
$\vcenter{\hbox{\includegraphics[width=4cm]{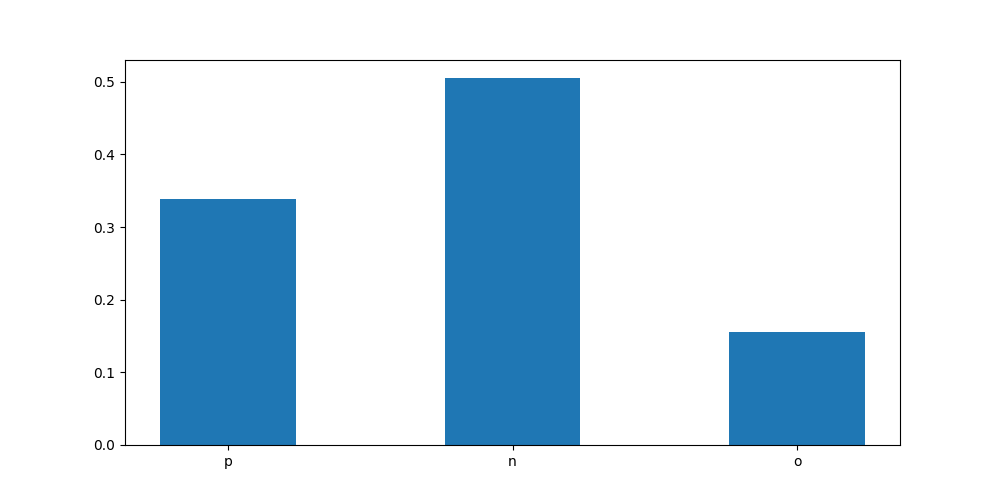}}}$
\\[-0.3em]
&
$\scriptstyle 0.3525\ket{p} + 0.5298\ket{n} + 0.1177\ket{o}$
& &
$\scriptstyle 0.3392\ket{p} + 0.5047\ket{n} + 0.1561\ket{o}$
\\[+1em]
\begin{tabular}{c} \textbf{ after } \\[-0.4em]
   \textbf{ optimistic } \\[-0.4em]
   \textbf{ preparation } \end{tabular} &
$\vcenter{\hbox{\includegraphics[width=4cm]{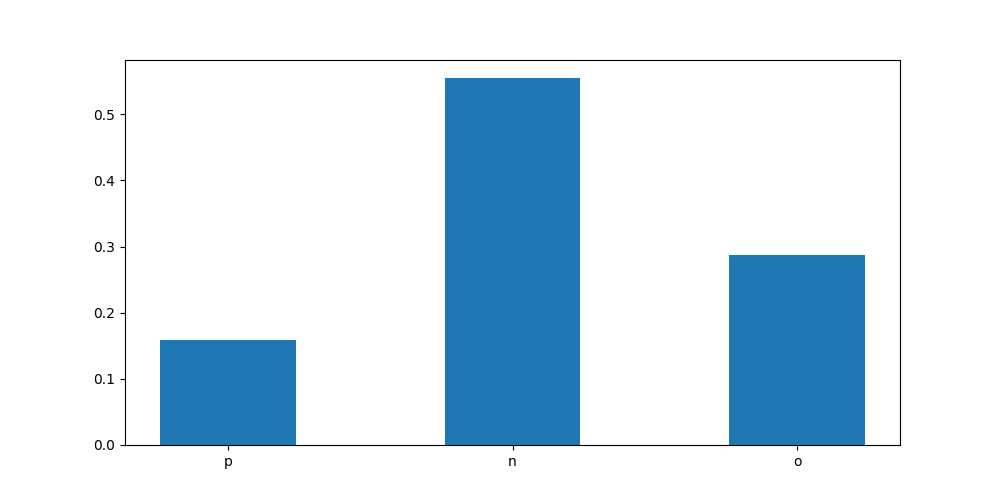}}}$
& &
$\vcenter{\hbox{\includegraphics[width=4cm]{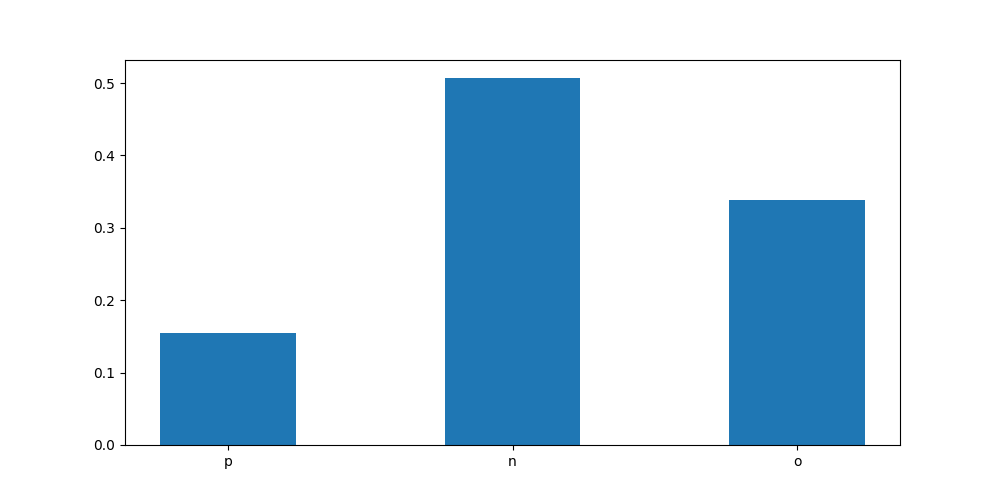}}}$
\\[-0.3em]
&
$\scriptstyle 0.1581\ket{p} + 0.5544\ket{n} + 0.2875\ket{o}$
& &
$\scriptstyle 0.1549\ket{p} + 0.5066\ket{n} + 0.3384\ket{o}$
\end{tabular}
\end{center}
\caption{Managing expectations before mood update with bad marks,
  after pessimistic preparation in the top row (with $r$) and after
  optimistic preparation (with the negation $r^\bot$) in the bottom
  row.}
\label{MoodPrepareUpdateFig}
\end{figure}

\noindent The resulting updated moods are described in the top row of
Figure~\ref{MoodPrepareUpdateFig}. Interestingly, if you try to
prepare for bad marks via a `pessimistic' predicate $r$, as in the
above bullet points, the resulting mood is more pessimistic than
without the preparation with $r$. In order to reduce the negative
impact of expected bad marks it works better to prepare positively, so
with the negation $r^{\bot}$ instead of with $r$, in the above two
points. This follows common wisdom: bracing for impact works better by
cheering up, since the bad news then hits less hard.


Overall we see that although there are considerable mathematical
differences between Pearl's and Jeffrey's update mechanism --- between
learning from what's right and learning from what's wrong --- they
react in the same directions to changes of focus and preparation.  In
this example we have described focus and preparation separately, but
of course they can be combined.

\section{Concluding remarks}\label{ConclusionSec}

This paper is a follow-up to earlier work of~\cite{Jacobs19c}, where
mathematically precise formulations were introduced for Pearl's and
Jeffrey's update rules. There, the distinction between the two rules
was described only in qualitative terms, namely as `improvement' (for
Pearl) versus `correction' (for Jeffrey). Here, this qualitative
characterisation is turned into a mathematical characterisation:
Pearl's rule increases validity, whereas Jeffrey's rule decreases
divergence.  The proof of the latter fact is the main technical
achievement of this paper.

The two update rules of Pearl and Jeffrey have been placed in the
setting of predictive coding theory. It remains an open question
whether these two update mechanisms can be distinguished empirically
in neuroscience.

\subsubsection*{Acknowledgements} Thanks are due to Harald Woracek and 
Ana Sokolova for pointing out the relevance of the
reference~\cite{FriedlandK75} for the proof of
Proposition~\ref{WeigthedUpdateProp} and for helpful subsequent
discussions.


\appendix
\section{Appendix}

This appendix contains a proof of the main result of this paper
(Theorem~\ref{JeffreyRuleThm}). The proof is extracted
from~\cite{FriedlandK75} and is specialised here to a conditional
expectations matrix. The original proof is formulated more generally.
We use some basic facts from linear algebra, esp.\ about non-negative
matrices, see \cite{Minc88} for background information. The proof also
uses Gelfand's spectral radius formula~\cite{Gelfand41}. This is a
``mathematical bazooka'', with a non-trivial proof, using linear
analysis. It is an open question if there is an easier proof for
Theorem~\ref{JeffreyRuleThm} in this paper.


We recall that for a square matrix $A$ the \emph{spectral radius}
$\rho(A)$ is the maximum of the absolute values of its eigenvalues:
\[ \begin{array}{rcl}
\rho(A)
& = &
\max\Big\{\, |\lambda|\;\big|\,\lambda \mbox{ is an eigenvalue of }A\,\Big\}.
\end{array} \]

\noindent We shall make use the following result. The first point is
known as Gelfand's formula, originally from~\cite{Gelfand41}. The
proof is non-trivial and is skipped here; for details see
\eg~\cite[Appendix~10]{Lax07}. For convenience we include short
(standard) proofs of the other two points.

\begin{theorem}
\label{SpectralRadiusThm}
Let $A$ be a (finite) square matrix, and $\|-\|$ be a matrix norm.
\begin{enumerate}
\item \label{SpectralRadiusThmGelfand} The spectral radius satisfies:
\[ \begin{array}{rcl}
\rho(A)
& = &
\lim\limits_{n\rightarrow\infty}\, \big\|\,A^{n}\,\big\|^{\nicefrac{1}{n}}.
\end{array} \]

\item \label{SpectralRadiusThmStochastic} Here we shall use the
  $1$-norm $\|\,A\,\|_{1} = \max_{j} \sum_{i} |A_{ij}|$. It
  yields that $\rho(A) = 1$ for each stochastic matrix
  $A$.

\item \label{SpectralRadiusThmBound} Let square matrix $A$ now be
  non-negative, that is, satisfy $A_{ij} \geq 0$, and let $x$ be a
  positive vector, so each $x_{i} > 0$. If $A x \leq r\cdot x$
  with $r>0$, then $\rho(A) \leq r$.
\end{enumerate}
\end{theorem}

\begin{myproof}
As mentioned, we skip the proof of the Gelfand's formula. If $A$ is
stochastic, one gets:
\[ \begin{array}{rcccccccl}
\|\,A\,\|_{1} 
& = &
\max_{j} \sum_{i} |A_{ij}|
& = &
\max_{j} \sum_{i} A_{ij}
& = &
\max_{j} 1
& = &
1.
\end{array} \]

\noindent Stochastic matrices are closed under matrix multiplication,
so  $\|\,A^{n}\,\|_{1} = 1$ for each $n$. Hence $\rho(A) = 1$ via
Gelfand's formula.


We next show how the third point can be obtained from the first one,
as in~\cite[Cor.~8.2.2]{Karlin59}.  By assumption, each entry $x_i$ in
the (finite) vector $x$ is positive.  Let's write $x_{-}$ for the
least one and $x_{+}$ for the greatest one.  Then $0 < x_{-} \leq
x_{i} \leq x_{+}$ for each $i$. For each $n$ we have:
\[ \begin{array}{rcl}
\big\|\,A^{n}\,\big\|_{1}\cdot x_{-}
& = &
\max_{j} \sum_{i}\, \big|\,\big(A^{n}\big)_{ij}\,\big|\cdot x_{-}
\\
& \leq &
\max_{j} \sum_{i}\, \big(A^{n}\big)_{ij}\cdot x_{i}
\\
& = &
\max_{j}\, \big(A^{n} x\big)_{j}
\\
& \leq &
\max_{j}\, \big(r^{n} \cdot x\big)_{j}
\\
& \leq &
r^{n} \cdot x_{+}.
\end{array} \]

\noindent Hence:
\[ \begin{array}{rcccl}
\big\|\,A^{n}\,\big\|_{1}^{\,\nicefrac{1}{n}}
& \leq &
\left(r^{n} \cdot \displaystyle\frac{x_{+}}{x_{-}}\right)^{\nicefrac{1}{n}}
& = &
r \cdot \left(\displaystyle\frac{x_{+}}{x_{-}}\right)^{\nicefrac{1}{n}}.
\end{array} \]

\noindent Thus, by Gelfand's formula, in the first point,
\[ \begin{array}[b]{rcl}
\rho(A)
& = &
\lim\limits_{n\rightarrow\infty}\, \big\|\,A^{n}\,\big\|^{\nicefrac{1}{n}}
\\
& \leq &
\lim\limits_{n\rightarrow\infty}\,r \cdot 
   \left(\displaystyle\frac{x_{+}}{x_{-}}\right)^{\nicefrac{1}{n}}
\hspace*{\arraycolsep}=\hspace*{\arraycolsep}
r \cdot \lim\limits_{n\rightarrow\infty}\,
   \left(\displaystyle\frac{x_{+}}{x_{-}}\right)^{\nicefrac{1}{n}}
\hspace*{\arraycolsep}=\hspace*{\arraycolsep}
r \cdot 1
\hspace*{\arraycolsep}=\hspace*{\arraycolsep}
r.
\end{array} \eqno{\QEDbox} \]
\end{myproof}

For the remainder of this appendix the setting is as follows. Let
$\omega\in\Dst(X)$ be a fixed state, with an $n$-test $p_{1}, \ldots,
p_{n}\in \Pred(X)$ so that $\sum_{i}p_{i} = \one$, by definition (of
test). We shall assume that the validities $v_{i} = \omega\models
p_{i}$ are non-zero. Notice that $\sum_{i} v_{i} = 1$. We organise
these validities in a vector $v$ and in diagonal matrix $V$:
\[ \begin{array}{rccclcrcl}
v
& = &
\left(\begin{array}{c}
v_{1} 
\\[-0.6em]
\vdots
\\[-0.6em]
v_{n}
\end{array}\right)
& = &
\left(\begin{array}{c}
\omega\models p_{1} 
\\[-0.6em]
\vdots
\\[-0.6em]
\omega\models p_{n}
\end{array}\right)
& \qquad\qquad &
V
& = &
\left(\begin{array}{ccc}
v_{1} & & 0 
\\[-0.6em]
& \ddots
\\[-0.6em]
0 & & v_{n}
\end{array}\right).
\end{array} \]

\noindent In addition, we use two $n\times n$ (real, non-negative)
matrices $B$ and $C$ given by:
\[ \begin{array}{rcl}
B_{ij}
& = &
\displaystyle\frac{\omega\models p_{i}\andthen p_{j}}
   {(\omega\models p_{i}) \cdot (\omega\models p_{j})}
\\[+1em]
C_{ij}
& = &
\omega|_{p_j} \models p_{i}
\hspace*{\arraycolsep}\,\smash{\stackrel{\eqref{BayesEqn}}{=}}\,\hspace*{\arraycolsep}
\displaystyle\frac{\omega\models p_{i}\andthen p_{j}}
   {\omega\models p_{j}}
\hspace*{\arraycolsep}=\hspace*{\arraycolsep}
(\omega\models p_{i})\cdot B_{ij}.
\end{array} \]

\noindent The next series of facts is extracted
from~\cite{FriedlandK75}.

\begin{lemma}
\label{WeigthedUpdateProofMatrixBasicsLem}
The above matrices $B$ and $C$ satisfy the following properties.
\begin{enumerate}
\item \label{WeigthedUpdateProofMatrixBasicsLemB} The matrix $B$ is
  non-negative and symmetric, and satisfies $B\,v = \one$.  Moreover,
  $B$ is positive definite, so that its eigenvalues are positive
  reals.

\auxproof{
If $B\,z = \lambda z$, for $z \neq 0$, then
$0 \leq z^{T}B\,z = z^{T}\lambda z
= \lambda\|z\|^{2}$. But then $\lambda\geq 0$.

}

\item \label{WeigthedUpdateProofMatrixBasicsLemBinv} As a result, the
  inverse $B^{-1}$ and square root $B^{\nicefrac{1}{2}}$ exist --- and
  $B^{-\nicefrac{1}{2}}$ too.

\item \label{WeigthedUpdateProofMatrixBasicsLemC} The matrix $C$ of
  conditional expectations is stochastic and thus its spectral radius
  $\rho(C)$ equals $1$, by
  Theorem~\ref{SpectralRadiusThm}~\eqref{SpectralRadiusThmStochastic}. Moreover,
  $C$ satisfies $C\,v = v$ and $C = V\cdot B$.

\item \label{WeigthedUpdateProofMatrixBasicsLemD} For an $n\times n$
  real matrix $D$, $\rho(DC) =
  \rho\big(B^{\nicefrac{1}{2}}DVB^{\nicefrac{1}{2}}\big)$.

\item \label{WeigthedUpdateProofMatrixBasicsLemIneq} Assume now that
  $D$ is a diagonal matrix with numbers $d_{1}, \ldots, d_{n} \geq 0$
  on its diagonal. Then:
\[ \begin{array}{rcl}
\sum_{i} d_{i}\cdot v_{i}
& \leq &
\rho(DC).
\end{array} \]
\end{enumerate}
\end{lemma}

\begin{myproof}
\begin{enumerate}
\item Clearly, $B_{ij} = B_{ji}$. Further,
\[ \hspace*{-1em}\begin{array}{rcccccccccl}
\big(B\,v\big)_{i}
& = &
\displaystyle \sum_{j} \frac{\omega\models p_{i}\andthen p_{j}}
   {(\omega\models p_{i}) \cdot (\omega\models p_{j})} \cdot
   (\omega\models p_{j})
& = &
\displaystyle \frac{\omega\models p_{i}\andthen (\sum_{j} p_{j})}
   {\omega\models p_{i}}
& = &
\displaystyle\frac{\omega\models p_{i}\andthen \one}{\omega\models p_{i}}
& = &
\displaystyle\frac{\omega\models p_{i}}{\omega\models p_{i}}
& = &
1.
\end{array} \]

\noindent The matrix $B$ is positive definite since for a
non-zero vector $z = (z_{i})$ of reals:
\[ \begin{array}{rcll}
z^{T}B\,z
& = &
\displaystyle \sum_{i,j} z_{i} \cdot \frac{\omega\models p_{i}\andthen p_{j}}
   {(\omega\models p_{i}) \cdot (\omega\models p_{j})} \cdot z_{j}
\\[+0.5em]
& = &
\omega \models 
   \left(\sum_{i} \frac{z_i}{v_i}\cdot p_{i}\right) \andthen
   \left(\sum_{j} \frac{z_j}{v_j}\cdot p_{j}\right)
\\
& = &
\omega \models q \andthen q & \mbox{for }
   q = \sum_{i} \frac{z_i}{v_i}\cdot p_{i}
\\
& > &
0.
\end{array} \]

\noindent We have a strict inequality $>$ here since $q \geq
\frac{z_1}{v_1}\cdot p_{1}$ and $\omega\models p_{1} > 0$, by
assumption; in fact this holds for each $p_i$. Thus:
\[ \begin{array}{rcccccl}
\omega \models q \andthen q
& \geq &
\displaystyle\frac{z_{1}^{2}}{v_{1}^{2}}\cdot (\omega \models p_{1}\andthen p_{1})
& \geq &
\displaystyle\frac{z_{1}^{2}}{v_{1}^{2}}\cdot (\omega \models p_{1})^{2}
& > &
0.
\end{array} \]

\noindent The last inequality follows from Theorem~\ref{UpdateThm} and
Bayes' rule~\eqref{BayesEqn}:
\[ \begin{array}{rccclcrcl}
\omega\models p
& \,\leq\, &
\omega|_{p} \models p
& = &
\displaystyle\frac{\omega\models p\andthen p}{\omega\models p}
& \qquad\mbox{so}\qquad &
\big(\omega\models p\big)^{2}
& \leq &
\omega\models p\andthen p.
\end{array} \]

\item The square root $B^{\nicefrac{1}{2}}$ and inverse $B^{-1}$ are
  obtained in the standard way via spectral decomposition $B =
  Q\Lambda Q^{T}$ where $\Lambda$ is the diagonal matrix of
  eigenvalues $\lambda_{i}>0$ and $Q$ is an orthogonal matrix (so
  $Q^{T} = Q^{-1}$). Then: $B^{\nicefrac{1}{2}} =
  Q\Lambda^{\nicefrac{1}{2}}Q^{T}$ where $\Lambda^{\nicefrac{1}{2}}$ has
  entries $\lambda_{i}^{\nicefrac{1}{2}}$. Similarly, $B^{-1} =
  Q\Lambda^{-1}Q^{T}$, and $B^{-\nicefrac{1}{2}} =
  Q\Lambda^{-\nicefrac{1}{2}}Q^{T}$.

\item It is easy to see that all $C$'s columns add up to one:
\[ \begin{array}{rcccccccl}
\sum_{i}C_{ij}
& = &
\sum_{i} \omega|_{p_j} \models p_{i}
& = &
\omega|_{p_j} \models \sum_{i} p_{i}
& = &
\omega|_{p_j} \models \one
& = &
1.
\end{array} \]

\noindent This makes $C$ stochastic, so that $\rho(C) = 1$. Next:
\[ \begin{array}{rcl}
\big(C\,v\big)_{i}
& = &
\sum_{j}\, (\omega|_{p_j} \models p_{i})\cdot (\omega\models p_{j})
\\
& = &
\sum_{j}\, \omega \models p_{i} \andthen p_{j}
   \qquad \mbox{by Bayes' rule~\eqref{BayesEqn}}
\\
& = &
\omega \models p_{i} \andthen (\sum_{j} p_{j})
\hspace*{\arraycolsep}=\hspace*{\arraycolsep}
\omega \models p_{i} \andthen \one
\hspace*{\arraycolsep}=\hspace*{\arraycolsep}
\omega \models p_{i} 
\hspace*{\arraycolsep}=\hspace*{\arraycolsep}
v_{i}.
\end{array} \]

\noindent Further, $\big(VB\big)_{ij} = v_{i}\cdot B_{ij} = C_{ij}$.

\item We show that $DC$ and $B^{\nicefrac{1}{2}}DVB^{\nicefrac{1}{2}}$
  have the same eigenvalues, which gives $\rho(DC) =
  \rho(B^{\nicefrac{1}{2}}DVB^{\nicefrac{1}{2}})$. First, let $DC\,z
  = \lambda z$. Take $z' =
  B^{\nicefrac{1}{2}}z$ one gets:
\[ \begin{array}{rcccccccccl}
B^{\nicefrac{1}{2}}DVB^{\nicefrac{1}{2}}z'
& = &
B^{\nicefrac{1}{2}}DVB^{\nicefrac{1}{2}}B^{\nicefrac{1}{2}}z
& = &
B^{\nicefrac{1}{2}}DCz
& = &
B^{\nicefrac{1}{2}}\lambda z
& = &
\lambda B^{\nicefrac{1}{2}}z
& = &
\lambda z'.
\end{array} \]

\noindent In the other direction, let
$B^{\nicefrac{1}{2}}DVB^{\nicefrac{1}{2}}w = \lambda w$. Now take $w' =
B^{-\nicefrac{1}{2}}w$ so that:
\[ \begin{array}{rcccccccl}
DCw'
& = &
B^{-\nicefrac{1}{2}}B^{\nicefrac{1}{2}}DVBB^{-\nicefrac{1}{2}}w
& = &
B^{-\nicefrac{1}{2}}\big(B^{\nicefrac{1}{2}}DVB^{\nicefrac{1}{2}}\big)w
& = &
B^{-\nicefrac{1}{2}}\lambda w
& = &
\lambda w'.
\end{array} \]

\auxproof{
Let $n\times n$ matrix $A$ have spectral radius $\rho(A)$. Then for
all non-zero vectors $x$, we have the following inequality for the
fraction of inner products:
\[ \begin{array}{rcl}
\displaystyle\frac{\big|\,(Ax,x)\,\big|}{(x,x)}
& \leq &
\rho(A)
\end{array} \]

\noindent To prove this, consider the eigen (spectral) decomposition
$A = Q\Lambda Q^{T}$, where $\Lambda$ is the diagonal matrix of
eigenvalues $\lambda_i$, and $Q$ is an orthogonal matrix (so $Q^{T} =
Q^{-1}$).  We first note:
\[ \begin{array}{rcccccccl}
\big|\,(\Lambda x,x)\,\big|
& = &
\big|\,\sum_{i} \lambda_{i}x_{i}^{2}\,\big|
& \leq &
\sum_{i} |\lambda_{i}|x_{i}^{2}
& \leq &
\sum_{i} \rho(A) x_{i}^{2}
& = &
\rho(A)(x,x).
\end{array} \]

\noindent But then:
\[ \begin{array}{rcccccl}
\displaystyle\frac{\big|\,(Ax,x)\,\big|}{(x,x)}
& = &
\displaystyle\frac{\big|\,(Q\Lambda Q^{T}x,x)\,\big|}{(QQ^{T}x,x)}
& = &
\displaystyle\frac{\big|\,(\Lambda Q^{T}x,Q^{T}x)\,\big|}{(Q^{T}x,Q^{T}x)}
& \leq &
\rho(A).
\end{array} \]

\noindent Here I am assuming that $Q$ is a real matrix, so that its
adjoint transpose $Q^{\dag} = \overline{Q}^{T}$ equals $Q^{T} =
Q^{-1}$.
}

\item We use the standard fact that for non-zero vectors $z$ one has:
\[ \begin{array}{rcl}
\displaystyle\frac{|\,(Az,z)\,|}{(z,z)}
& \leq &
\rho(A),
\end{array} \]

\noindent where $(-,-)$ is inner product. In particular,
\[ \begin{array}{rcccl}
\displaystyle\frac{|\,(B^{\nicefrac{1}{2}}DVB^{\nicefrac{1}{2}}z,z)\,|}{(z,z)}
& \leq &
\rho\big(B^{\nicefrac{1}{2}}DVB^{\nicefrac{1}{2}}\big)
& = &
\rho(DC).
\end{array} \]

\noindent We instantiate with $z = B^{\nicefrac{1}{2}}v$
and use that $VBv = Cv = v$ and
$Bv = \one$ in:
\[ \begin{array}[b]{rcl}
\rho(DC)
& \geq &
\displaystyle\frac{|\,(B^{\nicefrac{1}{2}}DVB^{\nicefrac{1}{2}}B^{\nicefrac{1}{2}}v,B^{\nicefrac{1}{2}}v)\,|}{(B^{\nicefrac{1}{2}}v,B^{\nicefrac{1}{2}}v)}
\hspace*{\arraycolsep}=\hspace*{\arraycolsep}
\displaystyle\frac{|\,(DVBv,B^{\nicefrac{1}{2}}B^{\nicefrac{1}{2}}v)\,|}{(v,B^{\nicefrac{1}{2}}B^{\nicefrac{1}{2}}v)}
\\[+1em]
& = &
\displaystyle\frac{|\,(Dv,Bv)\,|}{(v,Bv)}
\hspace*{\arraycolsep}=\hspace*{\arraycolsep}
\displaystyle\frac{|\,(Dv,\one)\,|}{(v,\one)}
\hspace*{\arraycolsep}=\hspace*{\arraycolsep}
\displaystyle\frac{|\,\sum_{i}d_{i}\cdot v_{i}\,|}{\sum_{i}v_{i}}
\hspace*{\arraycolsep}=\hspace*{\arraycolsep}
\textstyle\sum_{i}d_{i}\cdot v_{i}.
\end{array} \eqno{\QEDbox} \]
\end{enumerate}
\end{myproof}

\begin{proposition}
\label{WeigthedUpdateProp}
Let $\omega\in\Dst(X)$ be a state with predicates $p_{1}, \ldots,
p_{n}\in\Pred(X)$ forming a `test', so that $p_{1} + \cdots + p_{n}$
equal the constant 1 predicate $\one$. We assume $\omega\models p_{i}
\neq 0$, for each $i$. For all numbers $r_{1},\ldots,r_{n}\in (0,1]$
with $\sum_{i}r_{i} = 1$, one has:
\begin{equation}
\label{WeigthedUpdateEqn}
\begin{array}{rcl}
{\displaystyle \sum}_{i}\, \displaystyle\frac{r_{i} \cdot (\omega\models p_{i})}
   {\sum_{j} r_{j} \cdot (\omega|_{p_j}\models p_{i})}
& \;\leq\; &
1.
\end{array}
\end{equation}
\end{proposition}

\begin{myproof}
Consider a vector $r$ of non-zero numbers $r_{i} \in (0,1]$
  with $\sum_{i}r_{i} = 1$. We form a diagonal matrix $D$ with
  non-zero diagonal entries $d_{1},\ldots, d_{n}$ with:
\[ \begin{array}{rcccl}
d_{i}
& = &
\displaystyle\frac{r_{i}}{\big(C r\big)_{i}}
& = &
\displaystyle\frac{r_i}{\sum_{j} C_{ij}\cdot r_{j}}.
\end{array} \]

\noindent A crucial observation is that $r$ is an eigenvector
of the matrix $DC$, with eigenvalue $1$, since:
\[ \begin{array}{rcccccccl}
\big(DCr\big)_{i}
& = &
\sum_{j} \big(DC\big)_{ij}\cdot r_{j}
& = &
\sum_{j} d_{i} \cdot C_{ij} \cdot r_{j}
& = &
\displaystyle\frac{r_i}{\big(C r\big)_{i}} \cdot
   \textstyle\big(\sum_{j} C_{ij}\cdot r_{j}\big)
& = &
r_{i}.
\end{array} \]

\noindent
Theorem~\ref{SpectralRadiusThm}~\eqref{SpectralRadiusThmBound} now
yields $\rho(DC) = 1$.

By Lemma~\ref{WeigthedUpdateProofMatrixBasicsLem}~\eqref{WeigthedUpdateProofMatrixBasicsLemIneq} we get the required inequality in Proposition~\ref{WeigthedUpdateProp}:
\[ \begin{array}{rcccccl}
{\displaystyle\sum}_{i}\, \displaystyle\frac{r_{i}\cdot v_{i}}{\big(C r\big)_{i}}
& = &
\sum_{i} d_{i}\cdot v_{i}
& \leq &
\rho(DC)
& = &
1.
\end{array} \eqno{\QEDbox} \]
\end{myproof}

Finally we come to the proof of the main result.

\begin{myproof}[Of Theorem~\ref{JeffreyRuleThm}]
We reason as follows:
\[ \begin{array}{rcl}
\lefteqn{\KLD\Big(\tau, c \push \big(c_{\sigma}^{\dag} \push \tau\big)\Big)
   \;-\; \KLD\Big(\tau, c \push \sigma\Big)}
\\[+0.5em]
& = &
{\displaystyle\sum}_{y}\,\tau(y)\cdot\log\left(\displaystyle
   \frac{\tau(y)}{\big(c \push (c_{\sigma}^{\dag} \push \tau)\big)(y)}\right)
   \;-\;
   {\displaystyle\sum}_{y}\,\tau(y)\cdot\log\left(\displaystyle
   \frac{\tau(y)}{\big(c \push \sigma\big)(y)}\right)
\\[+1em]
& = &
{\displaystyle\sum}_{y}\,\tau(y)\cdot\log\left(\displaystyle
   \frac{\tau(y)}{\big(c \push (c_{\sigma}^{\dag} \push \tau)\big)(y)}
   \cdot\frac{\big(c \push \sigma\big)(y)}{\tau(y)}\right)
\\[+1em]
& \leq &
\log\left({\displaystyle\sum}_{y}\,\displaystyle
   \frac{\tau(y)\cdot \big(c \push \sigma\big)(y)}
        {\big(c \push (c_{\sigma}^{\dag} \push \tau)\big)(y)}\right)
\\[+1em]
& \leq &
\log\big(1\big) 
\hspace*{\arraycolsep}=\hspace*{\arraycolsep}
0.
\end{array} \]

\noindent The first inequality is an instance of Jensen's inequality.
The second one follows from Proposition~\ref{WeigthedUpdateProp}, via:
\[ \begin{array}{rcl}
{\displaystyle\sum}_{y}\, 
   \displaystyle\frac{\tau(y)\cdot \big(c \push \sigma\big)(y)}
        {\big(c \push (c_{\sigma}^{\dag} \push \tau)\big)(y)}
& = &
{\displaystyle\sum}_{y}\,\displaystyle
   \frac{\tau(y)\cdot (\sigma \models c \pull \indic{y})}
        {\sum_{x} (c_{\sigma}^{\dag} \push \tau)(x)\cdot c(x)(y)}
\\[+1.2em]
& = &
{\displaystyle\sum}_{y}\,\displaystyle
   \frac{\tau(y)\cdot (\sigma \models c \pull \indic{y})}
        {\sum_{x,z} \tau(z) \cdot c_{\sigma}^{\dag}(z)(x) \cdot (c \pull \indic{y})(x)}
\\[+1.2em]
& \smash{\stackrel{\eqref{DaggerEqn}}{=}} &
\displaystyle\sum_{y}
   \frac{\tau(y)\cdot (\sigma \models c \pull \indic{y})}
        {\sum_{x,z} \tau(z) \cdot 
         \frac{\sigma(x) \cdot (c \pull \indic{z})(x)}
              {\sigma \models c \pull \indic{z}} \cdot (c \pull \indic{y})(x)}
\\[+1.3em]
& = &
{\displaystyle\sum}_{y}\,\displaystyle
   \frac{\tau(y)\cdot (\sigma \models c \pull \indic{y})}
        {\sum_{z} \tau(z) \cdot 
        \frac{\sigma \models (c \pull \indic{z}) \andthen (c \pull \indic{y})}
             {\sigma \models c \pull \indic{z}}}
\\[+1.5em]
& \smash{\stackrel{\eqref{BayesEqn}}{=}} &
{\displaystyle\sum}_{y}\,\displaystyle
   \frac{\tau(y)\cdot (\sigma \models c \pull \indic{y})}
        {\sum_{z} \tau(z)\cdot(\sigma|_{c \pull \indic{z}} \models c \pull \indic{y})}
\\[+1em]
& \leq &
1, \qquad \mbox{by Proposition~\ref{WeigthedUpdateProp}.}
\end{array} \]

\noindent In the last line we apply
Proposition~\ref{WeigthedUpdateProp} with test $p_{i} \coloneqq c \pull
\indic{y_i}$, where $Y = \{y_{1}, \ldots, y_{n}\}$.  The point
predicates $\indic{y_i}$ form a test on $Y$. Predicate transformation
preserves tests. As assume in Theorem~\ref{JeffreyRuleThm}, $c \push
\sigma$ has full support, so that $\sigma \models p_{i} = \sigma
\models c \pull \indic{y_i} = (c \push \sigma)(y_{i})$ is non-zero for
each $i$. \QED
\end{myproof}

\end{document}